%% file: TimDepFano.tex
\documentclass[
english,
aps, pra,
onecolumn,
preprintnumbers,
amsmath, amssymb,
longbibliography
]{revtex4-2}
\usepackage{iftex}
\ifPDFTeX
\usepackage[utf8]{inputenc}   
\usepackage[T1]{fontenc}
\usepackage{lmodern}          
\fi

\usepackage[english]{babel}

\usepackage{amsmath,amssymb,mathtools}
\usepackage{bm}               

\usepackage{graphicx}
\usepackage{dcolumn}          
\usepackage[section]{placeins} 
\usepackage{needspace}        

\usepackage{xcolor}           

\usepackage[colorlinks=true, linkcolor=blue, citecolor=blue, urlcolor=blue]{hyperref}

\usepackage{microtype}

\usepackage{booktabs} 

\usepackage{amsthm}
\usepackage{amsfonts}
\usepackage{mathrsfs}
\usepackage{soul}
\usepackage{slashed}    
\usepackage{cleveref}
\begin{document}
	\title{Time-resolved Rydberg dynamics in molecules}
	
	\author{Erfan Saydanzad$^{1}$, Van-Hung Hoang$^{2}$, and Uwe Thumm$^{2}$}
	
	\affiliation{$^{1}$Department of Physics, Kennesaw State University, Marietta, Georgia 30060, USA}
	\affiliation{$^{2}$J. R. Macdonald Laboratory, Department of Physics, Kansas State University, Manhattan, Kansas 66506, USA}

	\date{\today}
	
	\begin{abstract}
		We develop a general theoretical framework for the ultrafast dynamics of coherently populated autoionizing Rydberg states in molecules, whose coherent decay results in asymmetric Fano line shapes produced by interference between resonant autoionization and direct ionization.
		The asymmetric line shapes form in the photoelectron continuum during
		the autoionization lifetime of the decreasing Rydberg population.
		By parameterizing autoionization resonances and solving the time-dependent Schr\"odinger equation, we model the buildup of the Fano profile together with the decline of the coherently excited Rydberg-state population.
		We compare our calculated
		photoelectron yields with the XUV-pump--IR-probe photoemission experiment of Biswas {\em et al.,} Phys. Rev. A {\bf 110}, 043106 (2024),
		in which neutral ground-state CO$_2$ molecules are excited by an XUV pulse to the Henning sharp and diffuse Rydberg series and subsequently ionized by the delayed IR pulse.
		Based on our numerically modeled photoelectron yields, we analyze two complementary processes:
		(i) photoemission into the $X^2\Pi_g$ continuum of residual CO$_2^+$ ions, which exposes interferences between direct photoionization and excited-state autoionization, and
		(ii) IR-induced photoemission into the CO$_2^+$[\(B^2\Sigma_u^+\)] continuum, which reveals quantum beats between Rydberg states.
		Our close-coupling model provides a versatile, extendable framework for tracking molecular Rydberg-state relaxation and, more generally, light-driven coherent dynamics in complex quantum systems.
	\end{abstract}
	
	\maketitle
	
	\section{Introduction}\label{sec:Introduction}

	A Fano resonance arises when excitation of a discrete autoionizing state interferes with direct photoelectron emission. The interference of these two pathways results in an asymmetric spectral profile~\cite{fano1961}.
	This mechanism occurs in systems ranging from plasmonic and photonic structures to semiconductor heterostructures~\cite{miroshnichenko2010,luk2010,xu2000scattering}. Its time evolution is especially revealing in photoionization, for which laser control of the Fano line shape~\cite{ott2013lorentz} and the buildup of a Fano resonance profile have been observed directly in gaseous helium~\cite{Kaldun2016} and numerically simulated~\cite{wickenhauser2005,mercouris2007,argenti2010,chu2010,argenti2013,desrier2018}. Compared to atoms,	
	molecules add dense Rydberg manifolds, multiple decay pathways, and several ionic continuum channels to this discrete--continuum interference~\cite{liao2017,plunkett2019,fidler2022,biswas2024}. In molecules, a pump pulse can typically prepare multiple autoionizing states coherently, while a delayed probe maps both their population decrease and relative phase into photoelectron signals.
	Interpreting these signals requires a framework in which direct ionization, resonant autoionization, and probe-induced ionization are represented as coherent amplitudes in one coupled set of equations, not as separate rates.
	
	XUV-pump--IR-probe photoelectron spectroscopy resolves electronic dynamics on attosecond to femtosecond time scales in targets ranging from atoms and molecules to nanostructures, where it allows, for example, the imaging of transient plasmonic near fields~\cite{Li2016,Saydanzad2017,Li2018PRL,Saydanzad2018}.
	Energy-resolved Fourier analysis of the photoemission dynamics is particularly useful, as it isolates the spectral components of the electron wave packet, allowing for the identification of quantum beats between spectral constituents of the photoelectron wave packets~\cite{Thumm2008,Plunkett2022Raman,fidler2022}.	In conventional photoelectron interferometry, however, the kinetic-energy resolution is limited by the probe-pulse bandwidth and the beat-frequency resolution by the accessible delay range. 
	In the Raman-interferometry scheme of Ref.~\cite{Plunkett2022Raman}, a pump pulse coherently excites a wave packet of autoionizing argon states, and a delayed probe pulse redistributes population among these states by two-photon Raman transitions. The change of the wave packet is detected later, through the electrons released by the slow natural autoionization of each state (deferred detection). These electrons form narrow lines whose widths are set by the natural autoionization widths and, in practice, by the spectrometer resolution, rather than by the probe bandwidth. Their yields oscillate with the pump--probe delay with frequencies corresponding to the level spacings. The scheme thus provides high temporal and spectral resolution simultaneously. This probe mechanism is distinct from the direct one-photon near-infrared (IR) ionization investigated in the present work.
	We here aim at a comprehensive representation of the time-dependent molecular electronic dynamics in the time and frequency domains. With this goal in mind, we derive coupled equations of motion for the amplitudes assigned to bound, autoionizing, and continuum molecular states. We solve these equations numerically and apply this model to examine recently measured extreme-ultraviolet (XUV)-pump--IR-probe photoelectron spectra of CO$_2$~\cite{biswas2024}, focusing on photoelectron emission into the $X^2\Pi_g$ and $B^2\Sigma_u^+$ continua of residual CO$_2^+$ cations in their electronic ground and second excited states. In this experiment, photoelectron energy and angular distributions were recorded by velocity-map imaging (VMI)~\cite{biswas2024}, a technique widely used to study light-induced ionization of molecules~\cite{Lam2022,Wangjam2021} and strong-field photoelectron emission from nanoparticles~\cite{Powell2022,Saydanzad2022,Saydanzad2023,Saydanzad2025}.
	
	Comparatively fast photoelectrons in the \(X^2\Pi_g\) continuum of CO$_2^+$ are formed by the coherent superposition of direct XUV ionization and resonant autoionization, while slow electrons emitted into the CO$_2^+$[\(B^2\Sigma_u^+\)] continuum are due solely to IR ionization of the surviving Rydberg wave packet. We refer to these pathways throughout as the autoionization (\(X\)) channel and the IR-induced (\(B\)) channel, respectively. Their combined time- and Fourier-domain (frequency) response provides complementary information: the $X$ channel encodes the phase between direct and resonant emission, oscillating according to the detuning of the photoelectron energy from each Rydberg level, whereas the $B$ channel encodes the relative phase between pairs of Rydberg states and oscillates at their energy difference. In this context, ``energy'' refers to the photoelectron kinetic energy and ``frequency'' to the Fourier conjugate of the pump--probe delay. The two channels also differ in their time scales: \(B\)-channel electrons are released within the femtosecond probe window and resolve the wave-packet dynamics in time, whereas autoionization electrons of the \(X\) channel are released over the autoionization lifetimes and resolve narrow resonance structures in energy. Unlike in the deferred detection of Ref.~\cite{Plunkett2022Raman}, these structures are superimposed on direct ionization, and their delay dependence is dominated by direct--autoionization interference rather than by Rydberg--Rydberg beats (Sec.~\ref{sec:systematic_comparison}). Our purpose is to establish how the Rydberg-state energies and lifetimes are encoded in the photoelectron energy and quantum-beat frequency of each channel, and to determine which features in the photoelectron spectrum are robust against the deliberately simplified modeling of the molecular continua.
	
	This paper is organized as follows. Section~\ref{sec:theory} introduces the equations of motion for photoemission along different ionization paths. Full derivations are given in the Supplementary Material (SM). In the SM we develop the general close-coupling model and equations of motion, while a specialized application of this numerical model to CO$_2$ is discussed in the main text.
	In Sec.~\ref{sec:results} we present numerical results for the time evolution of photoelectron wave packets produced by autoionization and IR ionization of Rydberg states. We compare our calculated photoelectron yields with recently measured experimental data of Biswas {\em et al.}~\cite{biswas2024} and derive quantum beat spectra~\cite{Thumm2008} by a windowed Fourier transformation over the experimental XUV--IR delay range. Section~\ref{sec:conclusion} contains our conclusions and an assessment of the limitations of our model. Unless noted otherwise, atomic units are used throughout this work.

	\section{Theory}\label{sec:theory}

	\FloatBarrier
	\subsection{Channel Basis and Hamiltonian}\label{sec:main_channel_basis}
	
	The formalism developed in this subsection is general, in the sense that it applies to any molecule in which a series of Rydberg states is embedded in one or more ionization continua, allowing the distinction of  {\em ionization continua} (i.e.,{\em ionization channels}) by the electronic state of the singly charged residual molecular ion.
	To model the electronic dynamics of a molecule in the presence of an XUV pump and a time-delayed IR probe pulse, we write the time-dependent Schr\"odinger equation as
	\begin{equation}\label{eq_m:TimeDep_SE}
		i\frac{\partial}{\partial t} |\Psi(t)\rangle = [H_{\mathrm{ch}}+V_{\mathrm{CI}}+D(t)]\,|\Psi(t)\rangle .
	\end{equation}
	Here \(H_0=K_N+H_e=H_{\mathrm{ch}}+V_{\mathrm{CI}}\) is the field-free molecular Hamiltonian, partitioned into the channel-diagonal operator \(H_{\mathrm{ch}}=K_N+\sum_aP_aH_eP_a\) and channel-off-diagonal configuration interaction \(V_{\mathrm{CI}}=\sum_{a\neq b}P_aH_eP_b\). \(D(t)\) is the interaction with the laser pulses, detailed below. The symbols \(K_N\), \(H_e\), and \(P_a\) denote the nuclear kinetic-energy operator, electronic Hamiltonian, and orthogonal projection onto electronic channel \(a\), respectively. The complete Born--Oppenheimer construction of the eigenstates of \(H_{\mathrm{ch}}\), including their vibrational quantum numbers and energies, is given in Sec.~\ref{sec:sup_theory} of the SM. The interaction \(V_{\mathrm{CI}}\) couples Rydberg states to energetically open continua and thereby causes autoionization. We expand the molecular wave function in the resulting channel basis as
	\begin{align}\label{eq:Psi_t}
		|\Psi(t)\rangle
		=&\, c_g(t)\, e^{-iE_g t} |g\rangle
		+ \sum_{n} c_{\mathcal{R}_n}(t)\, e^{-iE_{\mathcal{R}_n} t} |\mathcal{R}_n\rangle \nonumber\\
		&+ \sum_{\mathcal{C},\alpha} \int d\varepsilon_{\alpha}\,
		c_{\mathcal{C}}(t,\varepsilon_{\alpha})\, e^{-i(E_{\mathcal{C}}+\varepsilon_{\alpha})t} |\mathcal{C},\varepsilon_{\alpha}\rangle
		+ \sum_{\mathcal{N},\beta} \int d\varepsilon_{\beta}\,
		c_{\mathcal{N}}(t,\varepsilon_{\beta})\, e^{-i(E_{\mathcal{N}}+\varepsilon_{\beta})t} |\mathcal{N},\varepsilon_{\beta}\rangle .
	\end{align}
	States and energies in Eq.~\eqref{eq:Psi_t} are related to the Born--Oppenheimer states of Sec.~\ref{sec:sup_theory} by
	\begin{align*}
		|g\rangle
		&\equiv|\Psi_{g,\bm\nu_g}\rangle,
		& E_g&\equiv E_{g,\bm\nu_g},\\
		|\mathcal R_n\rangle
		&\equiv|\Psi_{n,\bm\nu_n}\rangle,
		& E_{\mathcal R_n}&\equiv E_{n,\bm\nu_n},\\
		|\mathcal C,\varepsilon_\alpha\rangle
		&\equiv|\Psi_{c,\bm\nu_c}\rangle\otimes|\varepsilon_\alpha\rangle,
		& E_{\mathcal C}&\equiv E_{c,\bm\nu_c},\\
		|\mathcal N,\varepsilon_\beta\rangle
		&\equiv|\Psi_{d,\bm\nu_d}\rangle\otimes|\varepsilon_\beta\rangle,
		& E_{\mathcal N}&\equiv E_{d,\bm\nu_d}.
	\end{align*}
	Here \(g\) labels the neutral ground electronic state, \(n\)  a Rydberg electronic state, and \(c\) and \(d\) the residual-ion electronic states in the continuum channels that are coupled to the Rydberg manifold by \(V_{\mathrm{CI}}\). \(\bm\nu_i\) denotes the vibrational quantum numbers associated with electronic channel \(i\). Each calligraphic label stands for a pair of electronic and vibrational labels, \(\mathcal R_n\equiv(n,\bm\nu_n)\), \(\mathcal C\equiv(c,\bm\nu_c)\), and \(\mathcal N\equiv(d,\bm\nu_d)\). In our  numerical model, each electronic state is represented by a single vibrational level. The general case, in which the sums also run over all vibrational levels, is given in Sec.~\ref{sec:sup_pulses} of the SM. The indices \(\alpha\) and \(\beta\) label photoelectron partial waves.
	
	The total energies of \(|\mathcal C,\varepsilon_\alpha\rangle\) and \(|\mathcal N,\varepsilon_\beta\rangle\) are \(E_{\mathcal C}+\varepsilon_\alpha\) and \(E_{\mathcal N}+\varepsilon_\beta\), where \(E_{\mathcal C}\) and \(E_{\mathcal N}\) are the energies of the residual-ion states and \(\varepsilon_\alpha\) and \(\varepsilon_\beta\) are the asymptotic photoelectron kinetic energies. The electron states are energy normalized, \(\langle\varepsilon|\varepsilon'\rangle=\delta(\varepsilon-\varepsilon')\). Residual-ion states belonging to different electronic channels are orthogonal, \(\langle\mathcal C,\varepsilon_\alpha|\mathcal N,\varepsilon_\beta\rangle=0\). The coefficients \(c_g\) and \(c_{\mathcal R_n}\) are probability amplitudes of the bound states, whereas \(c_{\mathcal C}\) and \(c_{\mathcal N}\) are probability-amplitude densities. \(|c_{\mathcal C}(t,\varepsilon_\alpha)|^2\) and \(|c_{\mathcal N}(t,\varepsilon_\beta)|^2\) are thus populations per unit photoelectron energy.
	Assuming the expansion in Eq.~\eqref{eq:Psi_t} is exact (not truncated), the populations
	\(S_g(t)=|c_g(t)|^2\),
	\(S_{\mathcal{R}_n}(t)=|c_{\mathcal{R}_n}(t)|^2\),
	\(S_{\mathcal{C}}(t,\varepsilon_{\alpha})=|c_{\mathcal{C}}(t,\varepsilon_{\alpha})|^2\), and
	\(S_{\mathcal{N}}(t,\varepsilon_{\beta})=|c_{\mathcal{N}}(t,\varepsilon_{\beta})|^2\) satisfy the completeness relation
	\begin{equation}\label{eq:normalization}
		S_g(t)
		+ \sum_{n} S_{\mathcal{R}_n}(t)
		+ \sum_{\mathcal{C},\alpha}\int d\varepsilon_{\alpha}\, S_{\mathcal{C}}(t,\varepsilon_{\alpha})
		+ \sum_{\mathcal{N},\beta}\int d\varepsilon_{\beta}\, S_{\mathcal{N}}(t,\varepsilon_{\beta}) = 1.
	\end{equation}
	
\begin{figure}[!htbp]
		\centering
		\includegraphics[width=1.0\textwidth]{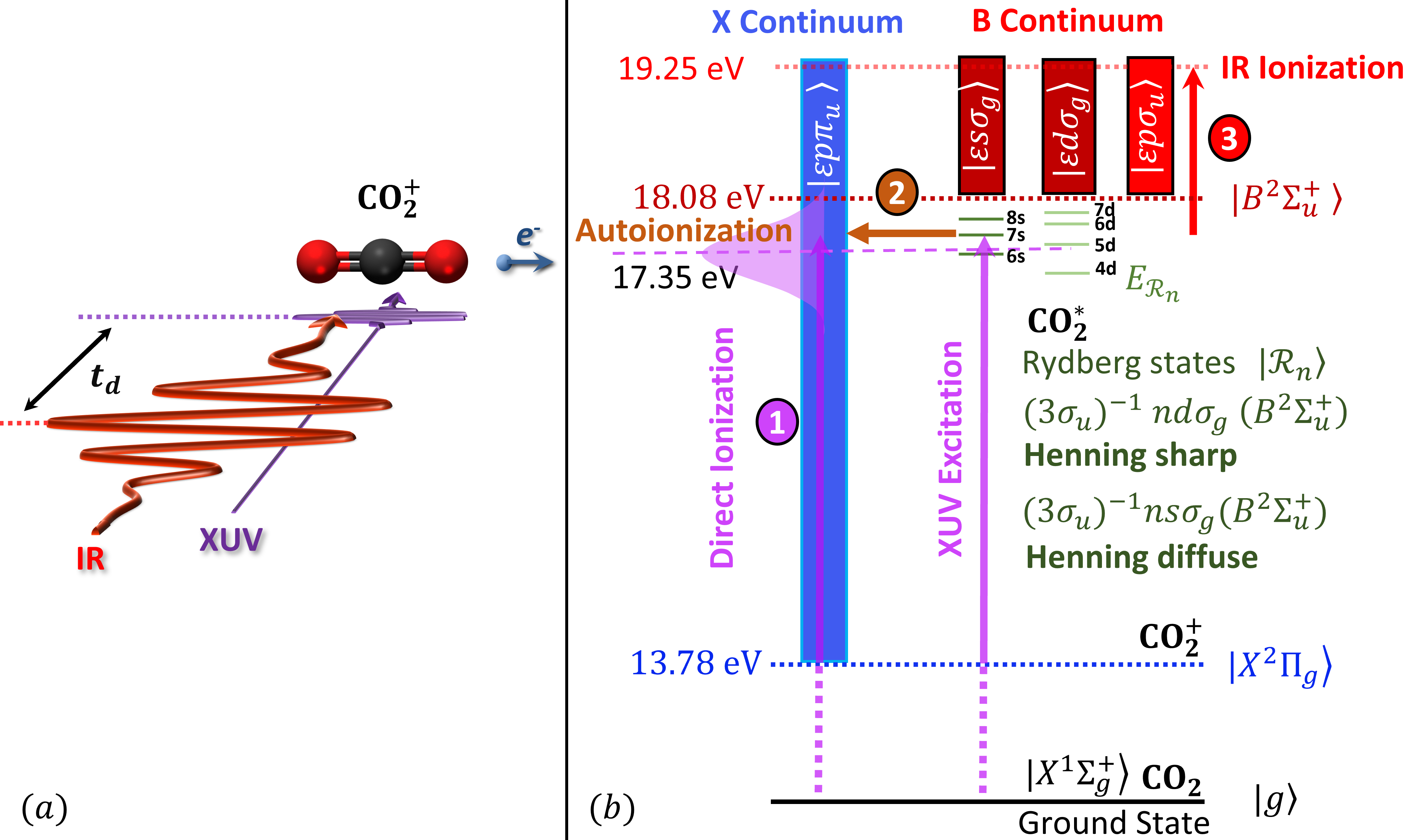}
		\caption{\textbf{XUV-pump/IR-probe excitation and ionization pathways and continuum channels for CO$_2$.}
			(a) XUV (purple) and IR (red) pulses separated by the pump--probe delay \(t_d\), ionizing CO$_2$ into CO$_2^+$ and a photoelectron \(e^-\).
			(b) Level diagram. From the ground state \(|g\rangle\), the XUV pulse either ionizes directly into the \(|X^2\Pi_g\rangle\) continuum of CO$_2^+$, with an ionization energy of 13.78~eV (pathway 1), or coherently excites the CO$_2^*$ Rydberg states \(|\mathcal{R}_n\rangle\) at energies \(E_{\mathcal R_n}\) of the \((3\sigma_u)^{-1}nd\sigma_g\) (Henning sharp) and \((3\sigma_u)^{-1}ns\sigma_g\) (Henning diffuse) series that converge to \(|B^2\Sigma_u^+\rangle\).
			The prepared \(|\mathcal{R}_n\rangle\) wave packet then either autoionizes into \(|X^2\Pi_g\rangle\) (pathway 2) or is ionized by the delayed IR pulse into \(|B^2\Sigma_u^+\rangle\) continuum with an ionization threshold of 18.08~eV (pathway 3).
			Above that threshold, the diagram marks the \(|\varepsilon p\sigma_u\rangle\) partial wave reachable by one-photon IR ionization of \(|\mathcal{R}_n\rangle\), and the \(|\varepsilon s\sigma_g\rangle\) and \(|\varepsilon d\sigma_g\rangle\) continua into which the \(ns\sigma_g\) and \(nd\sigma_g\) series converge. The states \(|\varepsilon p\pi_u\rangle\) represent photoelectron partial waves in the \(X^2\Pi_g\) continuum. The \(X\) continuum (pathways 1 and 2) and the \(B\) continuum (pathway 3) are distinguished by blue and red backgrounds and bold labels, matching the colors of their ionization thresholds.}
		\label{fig:Fig1}
	\end{figure}

	The time-dependent interaction
	\[
	D(t)=D_{\mathrm{XUV}}\,h_{\mathrm{XUV}}(t)\,e^{i(\omega_{\mathrm{XUV}}t+\phi_{\mathrm{XUV}})}
	+ D_{\mathrm{IR}}\,h_{\mathrm{IR}}(t-t_d)\,e^{i(\omega_{\mathrm{IR}}t+\phi_{\mathrm{IR}}-\omega_{\mathrm{IR}}t_d)} + \mathrm{h.c.},
	\]
	describes the interaction between the molecular dipole moment and external electric fields of the XUV-pump and IR-probe pulses,
	where h.c.\ denotes the Hermitian conjugate. \(D_{\mathrm{XUV}}\) and \(D_{\mathrm{IR}}\) are the corresponding dipole-interaction operators, \(h_{\mathrm{XUV}}\) and \(h_{\mathrm{IR}}\) the pulse envelopes, \(\omega_{\mathrm{XUV}}\) and \(\omega_{\mathrm{IR}}\) the carrier frequencies, and \(\phi_{\mathrm{XUV}}\) and \(\phi_{\mathrm{IR}}\) the carrier phases. \(t_d\) is the pump--probe delay [Fig.~\ref{fig:Fig1}(a)].
	Figure~\ref{fig:Fig1}(b) links these basis states and couplings to three pathways: direct XUV ionization into continuum channel(s) \(\mathcal{C}\), resonant excitation followed by autoionization into the same continuum (continua), and delayed IR ionization into the channel (channels) \(\mathcal{N}\). In the figure, continua are labeled by ionic term symbols, while thresholds keep the generic notation \(E_{\mathcal{C}}\) and \(E_{\mathcal{N}}\) of Eq.~\eqref{eq:Psi_t}.
	
	\FloatBarrier
	\subsection{Equations of Motion}\label{sec:main_eom}
	
	Projection of Eq.~\eqref{eq_m:TimeDep_SE} onto the channel basis, followed by the rotating-wave approximation (RWA) for the XUV and IR bound--continuum couplings,
	yields coupled equations for the amplitudes. Their derivations and general form are given in  Secs.~\ref{sec:sup_eom} and \ref{sec:sup_laplace_markov} of the SM. The derivation includes Laplace elimination of continuum amplitudes and separation of principal-value energy shifts from on-shell decay terms~\cite{kuerpick1997}.
	
	From this point onward, we specialize to CO\(_2\).  The generic channel labels of Sec.~\ref{sec:main_channel_basis} then take the concrete values shown in Fig.~\ref{fig:Fig1}(b). In addition to the RWA and Markov--Fano reduction, we make the following channel choices and numerical approximations:
	\begin{enumerate}
		\item[(i)] \emph{CO\(_2\) channel definition.} For the Henning Rydberg states considered here, spontaneous autoionization is observed predominantly into the \(X^2\Pi_g\) continuum~\cite{biswas2024}. We therefore retain the \(X^2\Pi_g\) continuum as the effective coupled autoionization channel and neglect possible coupling to the \(A^2\Pi_u\) continuum in the present numerical application.
		IR ionization populates the \(B^2\Sigma_u^+\) continuum. Thus,
		\[
		\mathcal{C}=X^2\Pi_g
		\qquad
		\mathcal{N}=B^2\Sigma_u^+,
		\]
		and we henceforth use the ionic labels \(X\) and \(B\) in place of \(\mathcal C\) and \(\mathcal N\).
		\item[(ii)] \emph{One vibrational level and one partial wave per channel.} To keep the numerical problem tractable, each electronic state is represented by a single Born--Oppenheimer vibrational level, and each ionic channel by a single photoelectron partial wave. The nuclear motion is frozen. For the \(B\) channel, the choice follows from the dipole selection rules of the \(D_{\infty h}\) point group of the linear, centrosymmetric triatomic CO\(_2\) molecule~\cite{SignorellMerkt1997}. A one-photon transition changes the total inversion parity, \(\pi_{\mathrm{total}}=\pi_{\mathrm{core}}(-1)^l\), where \(l\) is the photoelectron orbital angular momentum, while IR ionization leaves the \(B^2\Sigma_u^+\) core unchanged; the photoelectron parity \((-1)^l\) must therefore change.
		The gerade-parity \(ns\sigma_g\) and \(nd\sigma_g\) Rydberg electrons thus ionize only into ungerade, odd-\(l\) partial waves (\(\varepsilon p\), \(\varepsilon f\)). The anisotropic molecular potential mixes only partial waves of equal parity. Near threshold, the centrifugal barrier suppresses \(\varepsilon f\) relative to \(\varepsilon p\) partial waves.
		For the fixed parallel molecular orientation along the parallel polarizations of the pump and probe pulses, the projection of \(l\) on this axis, \(\lambda\), does not change, \(\Delta\lambda=0\), and the retained partial wave is \(\varepsilon p\sigma_u\).
		For the \(X\) channel, autoionization conserves the total \(\Sigma_u^+\) symmetry of the Henning states, and the \(X^2\Pi_g\) core level couples to a \(\varepsilon p\pi_u\) partial wave. Both choices agree with the measured photoelectron angular distributions~\cite{biswas2024} confirming that IR-ionization electrons are emitted along the polarization direction, and autoionization electrons perpendicular to it.
		\item[(iii)] \emph{Effective \(X\) continuum.} The \(X\)-channel waves coupled to \(|\mathcal R_n\rangle\) are replaced by one effective continuum,
		\begin{equation}\label{eq:main_effective_continuum}
			\langle X,\varepsilon_\alpha|V_{\mathrm{CI}}|\mathcal{R}_n\rangle
			=\sqrt{\frac{\Gamma_{nn}}{2\pi}}.
		\end{equation}
		Its coupling to each Rydberg state is real in the standing-wave phase convention of Sec.~\ref{sec:sup_eom} of the SM. It is assumed to be energy independent across the autoionization width \(\Gamma_{nn}=2\pi|\langle X,\varepsilon_\alpha|V_{\mathrm{CI}}|\mathcal R_n\rangle|^2\)~\cite{fano1961}.
		Continuum--continuum coupling is neglected, and the diagonal principal-value shifts \(F_{nn}\) are included in \(E_{\mathcal R_n}\) (Sec.~\ref{sec:sup_laplace_markov} of the SM).
		\item[(iv)] \emph{Coupling through the common \(X\) continuum.} Since all Rydberg states decay into the same effective continuum, they are coupled by the cross widths \(\Gamma_{nm}=\sqrt{\Gamma_{nn}\Gamma_{mm}}\), \(m\neq n\), which we retain. The off-diagonal level shifts \(F_{nm}\) are neglected.
		\item[(v)] \emph{IR couplings.} The IR pulse couples \(|\mathcal R_n\rangle\) only to the \(B\) continuum, in both directions, allowing ionization and coherent return during the probe pulse. Bound--bound IR couplings vanish, \(\langle\mathcal R_m|D_{\mathrm{IR}}|\mathcal R_n\rangle=0\), because the \(ns\sigma_g\) and \(nd\sigma_g\) states have the same parity. IR ionization of \(|g\rangle\) requires about nine IR photons and is neglected. IR ionization of \(|\mathcal R_n\rangle\) into the \(X\) continuum would require a change of the ionic core (\(B\to X\)), which the one-electron dipole operator does not produce, and is also neglected.
		
		\item[(vi)] \emph{Neglect of predissociation.} Henning sharp states can predissociate to neutral fragments [CO$(A^1\Pi)$ + O$(^1S)$]. This is a channel that competes with autoionization in this series~\cite{biswas2024,fidler2022}. This pathway is omitted here. Calculated sharp-series autoionization yields are thus upper bounds.
		
		\item[(vii)] \emph{Fixed molecular orientation.} In the experiment, the CO\(_2\) molecules are randomly oriented. We do not average over orientations and assume the molecular axis to be parallel to the common polarization direction of the pump and probe pulses. This orientation is favored in the experiment, since the Henning states are excited by a parallel transition~\cite{biswas2024}. Orientational averaging would change the relative amplitudes, but only weakly affects state energies, beat frequencies, or photoelectron energies.
	\end{enumerate}
	
	With these simplifications, the ground-state amplitude \(c_g(t)\) in Eq.~\eqref{eq:Psi_t}, coupled by the XUV-pulse electric field to Rydberg levels and the \(X\)-continuum channels, is given by
	\begin{align}\label{eq:c_g_dot}
		i\dot{c}_{g}(t)
		=& \sum_{n} c_{\mathcal{R}_n}(t)\,h_{\mathrm{XUV}}(t)\,e^{i\phi_{\mathrm{XUV}}}
		e^{-i(E_{\mathcal{R}_n}-E_g-\omega_{\mathrm{XUV}})t}
		\langle g|D_{\mathrm{XUV}}|\mathcal{R}_n\rangle \nonumber\\
		&+ \int d\varepsilon_{\alpha}\,
		c_X(t,\varepsilon_{\alpha})\,h_{\mathrm{XUV}}(t)\,e^{i\phi_{\mathrm{XUV}}}
		e^{-i(E_X+\varepsilon_{\alpha}-E_g-\omega_{\mathrm{XUV}})t}
		\langle g|D_{\mathrm{XUV}}|X,\varepsilon_{\alpha}\rangle .
	\end{align}
	Its squared modulus gives the neutral-ground-state population. Here and below, dots denote time derivatives. The two dipole matrix elements describe XUV excitation into a Rydberg state and direct XUV ionization, respectively. Similarly, the amplitudes of autoionizing Rydberg state $|\mathcal{R}_n\rangle$ obey
	\begin{align}\label{eq:final_Rydberg_EOM}
		i\,\dot c_{\mathcal{R}_n}(t) &+ \frac{i}{2}\sum_{m}\Gamma_{nm}\,
		e^{i(E_{\mathcal R_n}-E_{\mathcal R_m})t}\,c_{\mathcal{R}_m}(t)
		= \rho_n\,c_g(t)\,h_{\mathrm{XUV}}(t)\,e^{-i\phi_{\mathrm{XUV}}}\,
		e^{i(E_{\mathcal{R}_n}-E_g-\omega_{\mathrm{XUV}})t} \nonumber\\
		&\quad + \int d\varepsilon_{\beta}\,
		c_B(t,\varepsilon_{\beta})\,h_{\mathrm{IR}}(t-t_d)\,e^{i(\phi_{\mathrm{IR}}-\omega_{\mathrm{IR}}t_d)} \,
		e^{-i(E_B+\varepsilon_{\beta}-E_{\mathcal{R}_n}-\omega_{\mathrm{IR}})t}\,
		\langle \mathcal{R}_n|D_{\mathrm{IR}}|B,\varepsilon_{\beta}\rangle .
	\end{align}
	The sum on the left-hand side describes autoionization. The first term on the right-hand side of this equation corresponds to XUV excitation of the Rydberg state from the ground state with the pump strength
	\begin{equation}\label{eq:main_rho_n}
		\rho_n
		=(q_n-i)\sqrt{\frac{\pi\Gamma_{nn}}{2}}\,
		\langle g|D_{\mathrm{XUV}}|X,\varepsilon_\alpha\rangle
		\Big|_{\varepsilon_\alpha=E_{\mathcal R_n}-E_X},
	\end{equation}
	while the second term represents ionization of the Rydberg state $|\mathcal{R}_n\rangle$ to the \(B\) continuum, where the vertical bar denotes evaluation at the on-shell photoelectron energy \(\varepsilon_\alpha=E_{\mathcal R_n}-E_X\). In the channel convention we use, the asymmetry parameters \(q_n\) are real (cf., Sec.~\ref{sec:sup_laplace_markov} of the SM). Numerical values for \(q_n\), evaluated according to Eq.~\eqref{eq:q_scaling} below are given in Table~\ref{tab:Rydberg_States}.
	
	The \(X\)-continuum amplitudes are determined by direct ionization from the ground state (first term on the right-hand side) and by autoionization of the Rydberg states (second term),
	\begin{align}\label{eq:c_alpha_dot}
		i\dot{c}_X(t,\varepsilon_{\alpha})
		=&\, c_{g}(t)\,h_{\mathrm{XUV}}(t)\,e^{-i\phi_{\mathrm{XUV}}}
		e^{i(E_X+\varepsilon_{\alpha}-E_g-\omega_{\mathrm{XUV}})t}
		\langle g|D_{\mathrm{XUV}}|X,\varepsilon_{\alpha}\rangle \nonumber\\
		&+ \sum_{n} \sqrt{\frac{\Gamma_{nn}}{2\pi}} c_{\mathcal{R}_n}(t)
		e^{i(E_X+\varepsilon_{\alpha}-E_{\mathcal{R}_n})t}	 .
	\end{align}
	Interference between these two terms generates time-dependent asymmetric Fano line profiles.
	In contrast, the amplitudes of \(B\)-continuum states  are not directly coupled to the ground state  and fed only by IR ionization of the Rydberg wave packet,
	\begin{align}\label{eq:c_beta_dot}
		i\dot{c}_B(t,\varepsilon_{\beta})
		=&\, \sum_{n} c_{\mathcal{R}_n}(t)
		h_{\mathrm{IR}}(t-t_d)\,e^{-i(\phi_{\mathrm{IR}}-\omega_{\mathrm{IR}}t_d)}
		e^{-i(E_{\mathcal{R}_n}-E_B-\varepsilon_{\beta}+\omega_{\mathrm{IR}})t}
		\langle \mathcal{R}_n|D_{\mathrm{IR}}|B,\varepsilon_{\beta}\rangle .
	\end{align}
	The source and coupling terms in Eqs.~\eqref{eq:c_g_dot}--\eqref{eq:c_beta_dot} correspond to the three pathways labeled in Fig.~\ref{fig:Fig1}(b).

	\section{Numerical and Experimental Results}\label{sec:results}
	
	We scrutinized our theoretical model against a recent pump--probe photoelectron spectroscopy experiment on carbon dioxide (CO$_2$) molecules~\cite{biswas2024}. The pump--probe geometry and relevant energy-level scheme are shown in Fig.~\ref{fig:Fig1}. The pump and probe pulses are collinearly polarized. In this experiment, neutral ground-state CO$_2$ ($|X^1\Sigma_g^+ \rangle$) was excited to Rydberg states by the 11th harmonic of an attosecond pulse train. This harmonic was produced by focusing half of the 780 nm driving beam into a xenon gas cell. It is centered at 17.5~eV with a full width at half maximum (FWHM) bandwidth of 0.4~eV. The excitation populates two Rydberg-state progressions converging to the $B^2\Sigma_u^+$ ionic continuum at 18.08~eV, the Henning sharp series [$(3\sigma_u)^{-1}nd\sigma_g$] and the Henning diffuse series [$(3\sigma_u)^{-1}ns\sigma_g$]~\cite{Henning1932,TanakaJursaLeBlanc1960}. After excitation, the Rydberg states decay predominantly via spontaneous autoionization into the $X^2\Pi_g$ ionic continuum~\cite{biswas2024}. Possible coupling to the \(A^2\Pi_u\) continuum is omitted in the present model [Sec.~\ref{sec:main_eom}, item (i)]. A delayed 780~nm IR probe pulse with 50~fs FWHM, 1.59~eV photon energy, and a peak intensity at the interaction region of \(\sim1\)~TW/cm\(^2\) ionizes the surviving Rydberg population to the $B^2\Sigma_u^+$ state of CO$_2^+$.

	\FloatBarrier
	\subsection{State Parameters and Lifetime-Controlled Dynamics}\label{sec:dipole_scaling}
	
	Table~\ref{tab:Rydberg_States} collects the parameters for the 19 states used in the propagation. The energies are taken from the photoabsorption spectrum of Ref.~\cite{biswas2024}, adapted from Ref.~\cite{HuestisBerkowitz2011}. In the time-reversal-symmetric channel-phase convention of Fano theory~\cite{fano1961,FanoCooper1968}, all listed dipoles and asymmetry parameters are real. We define all dipole matrix elements and configuration-interaction couplings, and hence all \(q_n\) and \(\Gamma_{nm}\), as positive.
	
	\begin{table}[!htbp]
		\centering
		\renewcommand{\arraystretch}{1.15}
		\setlength{\tabcolsep}{5pt}
		\caption{\textbf{Parameters of the Henning sharp (\(nd\sigma_g\)) and diffuse (\(ns\sigma_g\)) states used in the CO$_2$ calculation.} Energies (relative to the ground state CO$_2$), lifetimes, dipole matrix elements (per unit field amplitude), and Fano asymmetry parameters \(q_n\).
			The IR dipole matrix elements are given relative to \(\langle\mathcal R_{5d}|D_{\mathrm{IR}}|B,\varepsilon_\beta\rangle\), which is set to 1 [Eq.~\eqref{eq:dipole_scaling_eq}]. Superscripts identify data sources.}
		\label{tab:Rydberg_States}
		\begin{tabular}{c r r c c r}
			\toprule
			\multicolumn{6}{c}{\textbf{Henning sharp series (\(nd\sigma_g\))}} \\
			\cmidrule(lr){1-6}
			State & Energy (eV)\textsuperscript{ b} & Lifetime (fs)
			&  \(\langle g|D_{\mathrm{XUV}}|\mathcal R_n\rangle\) (\(10^{-2}\))
			&  \(\langle\mathcal R_n|D_{\mathrm{IR}}|B,\varepsilon_\beta\rangle\)
			& \(q_n\) \\
			\midrule
			4d  &   17.2831 &    286.0\textsuperscript{ b} & 1.4461 & 1.338 & 1.25 \\
			5d  &   17.5395 &    372.6\textsuperscript{ a} & 2.0196 & 1.000 &  2.00 \\
			6d  &   17.7018 &   1457.5\textsuperscript{ a} & 1.7632 & 0.765 & 3.45 \\
			7d  &   17.8000 &   3378.5\textsuperscript{ a} & 1.3289 & 0.611 & 3.96 \\
			8d  &   17.8650 &   5021.2\textsuperscript{ d} & 1.1573 & 0.501 & 4.21 \\
			9d  &   17.9091 &   7085.1\textsuperscript{ d} & 0.8442 & 0.422 & 3.65 \\
			10d &   17.9407 &   9628.0\textsuperscript{ d} & 0.6948 & 0.362 & 3.50 \\
			11d &   17.9644 &  12735.7\textsuperscript{ d} & 0.6041 & 0.315 & 3.50 \\
			12d &   17.9825 &  16442.0\textsuperscript{ d} & 0.5316 & 0.277 & 3.50 \\
			\bottomrule
		\end{tabular}
		
		\vspace{0.8em}
		
		\begin{tabular}{c r r c c r}
			\toprule
			\multicolumn{6}{c}{\textbf{Henning diffuse series (\(ns\sigma_g\))}} \\
			\cmidrule(lr){1-6}
			State & Energy (eV)\textsuperscript{ b} & Lifetime (fs)
			&  \(\langle g|D_{\mathrm{XUV}}|\mathcal R_n\rangle\) (\(10^{-2}\))
			&  \(\langle\mathcal R_n|D_{\mathrm{IR}}|B,\varepsilon_\beta\rangle\)
			& \(q_n\) \\
			\midrule
			5s  &   17.0698 &     15.8\textsuperscript{ d} & 2.1489 & 1.726 & 0.44 \\
			6s  &   17.4579 &     33.0\textsuperscript{ c} & 2.0217 & 1.200 & 0.60 \\
			7s  &   17.6572 &     53.0\textsuperscript{ c} & 1.3713 & 0.898 & 0.51 \\
			8s  &   17.7725 &     94.0\textsuperscript{ c} & 1.1209 & 0.707 & 0.56 \\
			9s  &   17.8457 &    141.3\textsuperscript{ d} & 0.9836 & 0.577 & 0.60 \\
			10s &   17.8968 &    204.4\textsuperscript{ d} & 0.8240 & 0.480 & 0.60 \\
			11s &   17.9315 &    280.1\textsuperscript{ d} & 0.7291 & 0.410 & 0.63 \\
			12s &   17.9581 &    376.6\textsuperscript{ d} & 0.6287 & 0.353 & 0.63 \\
			13s &   17.9770 &    484.9\textsuperscript{ d} & 0.5542 & 0.311 & 0.63 \\
			14s &   17.9925 &    619.3\textsuperscript{ d} & 0.4904 & 0.276 & 0.63 \\
			\bottomrule
		\end{tabular}
		
		\begin{tabular}{l}
			\(^{\mathrm{a}}\) \emph{ab initio} from Ref.~\cite{fidler2022} \\
			\(^{\mathrm{b}}\) measured value from Ref.~\cite{biswas2024} \\
			\(^{\mathrm{c}}\) measured values from Table~I of Ref.~\cite{fidler2022} \\
			\(^{\mathrm{d}}\) according to Eq.~\eqref{eq:tau_scaling}
		\end{tabular}
		
	\end{table}

	The effective quantum number is \(n^*=[\mathcal E_{\mathrm{Ryd}}/(E_B-E_{\mathcal R_n})]^{1/2}\), where \(\mathcal E_{\mathrm{Ryd}}=13.606\)~eV and \(E_B=18.08\)~eV. Since \(\Gamma_{nn}\propto(n^*)^{-3}\), the lifetimes \(\tau_n=\Gamma_{nn}^{-1}\) of states without reported values are obtained from
	\begin{align}\label{eq:tau_scaling}
		\tau_{nd}
		&=\tau_{7d}
		\left(\frac{E_B-E_{\mathcal R_{7d}}}
		{E_B-E_{\mathcal R_{nd}}}\right)^{3/2},
		&& n=8,\ldots,12, \nonumber\\
		\tau_{ns}
		&=\tau_{8s}
		\left(\frac{E_B-E_{\mathcal R_{8s}}}
		{E_B-E_{\mathcal R_{ns}}}\right)^{3/2},
		&& n=5,9,\ldots,14.
	\end{align}
	The reported lifetimes are used directly. The highest reported member of each series anchors the extrapolation: \(\tau_{7d}=3378.5\)~fs for the sharp series and \(\tau_{8s}=94.0\)~fs for the diffuse series. For the 5d--7d states, these are the \emph{ab initio} values of Ref.~\cite{fidler2022}. For the 4d state, we use the measured value of Ref.~\cite{biswas2024}.
	
	For atomic Rydberg states near ionization threshold, bound--continuum matrix elements scale as \((n^*)^{-3/2}\)~\cite{BetheSalpeter,Hoogenraad1998}. We adopt this asymptotic scaling for the molecular couplings \(\langle\mathcal R_n|D_{\mathrm{IR}}|B,\varepsilon_\beta\rangle\),
	\begin{align}\label{eq:dipole_scaling_eq}
		\langle\mathcal R_{nd}|D_{\mathrm{IR}}|B,\varepsilon_\beta\rangle
		&=\langle\mathcal R_{5d}|D_{\mathrm{IR}}|B,\varepsilon_\beta\rangle
		\left(\frac{E_B-E_{\mathcal R_{nd}}}
		{E_B-E_{\mathcal R_{5d}}}\right)^{3/4},
		&& n=4,\ldots,12, \nonumber\\
		\langle\mathcal R_{ns}|D_{\mathrm{IR}}|B,\varepsilon_\beta\rangle
		&=\langle\mathcal R_{6s}|D_{\mathrm{IR}}|B,\varepsilon_\beta\rangle
		\left(\frac{E_B-E_{\mathcal R_{ns}}}
		{E_B-E_{\mathcal R_{6s}}}\right)^{3/4},
		&& n=5,\ldots,14.
	\end{align}
	We set \(\langle\mathcal R_{5d}|D_{\mathrm{IR}}|B,\varepsilon_\beta\rangle=1\) and \(\langle\mathcal R_{6s}|D_{\mathrm{IR}}|B,\varepsilon_\beta\rangle = 1.2\), fixing the overall scale and the diffuse-to-sharp coupling ratio.
	
\begin{figure}[htbp]
		\centering
		\includegraphics[width=\textwidth]{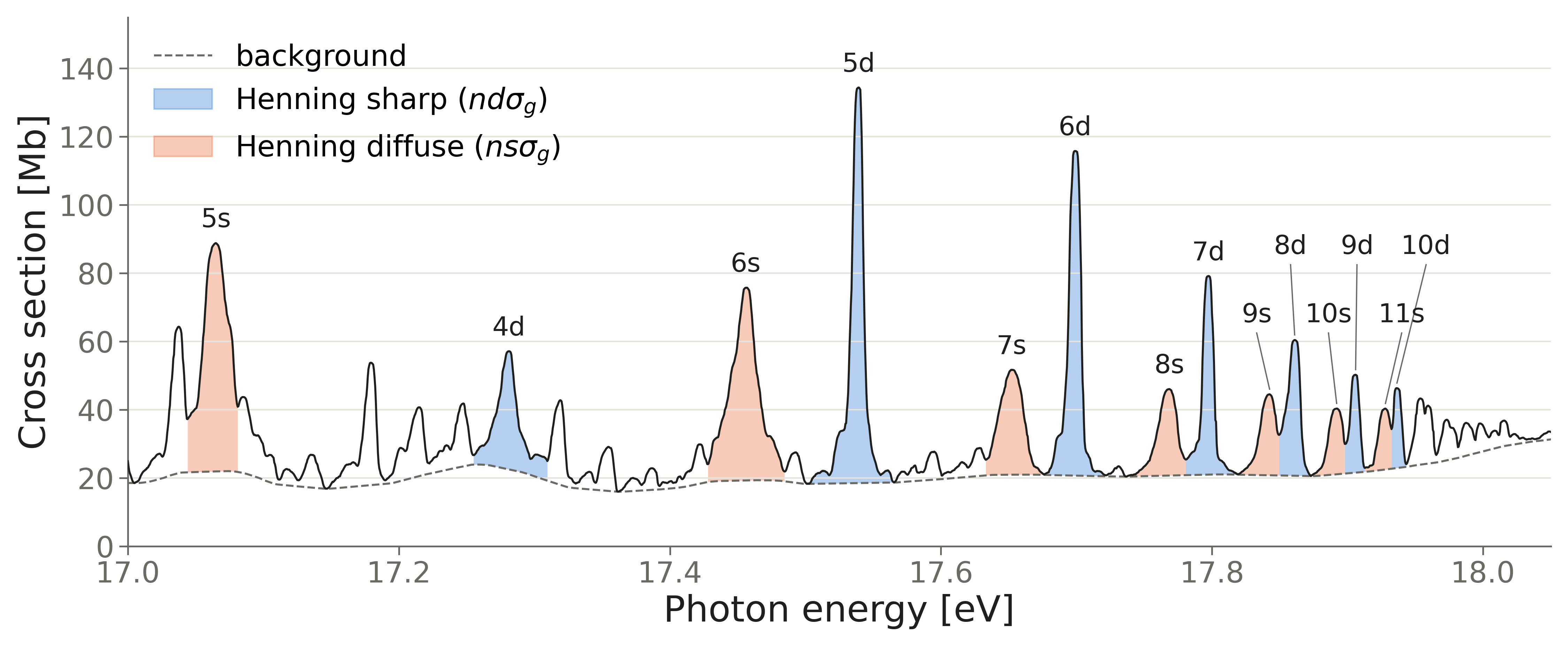}
		\caption{\textbf{Measured photoabsorption cross section of CO$_2$ between 17.0 and 18.05~eV}~\cite{HuestisBerkowitz2011}. The shaded areas above the smooth background (dashed) are the integrated line strengths of the Henning sharp (\(nd\sigma_g\), blue) and diffuse (\(ns\sigma_g\), orange) states, from which the relative XUV dipoles \(\langle g|D_{\mathrm{XUV}}|\mathcal R_n\rangle\) in Table~\ref{tab:Rydberg_States} are obtained.}
		\label{fig:photoabsorption}
	\end{figure}

	The relative values of \(\langle g|D_{\mathrm{XUV}}|\mathcal R_n\rangle\) are obtained from the integrated line areas of the photoabsorption spectrum evaluated by Huestis and Berkowitz~\cite{HuestisBerkowitz2011} (Fig.~\ref{fig:photoabsorption}), taken in digital form from the Leiden cross section database~\cite{Heays2017}. After subtraction of the smooth background, the area of each resonance is proportional to \(E\,|\langle g|D_{\mathrm{XUV}}|\mathcal R_n\rangle|^2\). The areas are independent of the spectral resolution, which broadens the sharp lines well beyond their natural widths. The lines are resolved up to the 10d and 11s states. For higher Rydberg states, the dipoles are extrapolated proportional to \((n^*)^{-3/2}\). The XUV-to-direct-ionization scale determines the asymmetry parameter according to
	\begin{equation}\label{eq:q_scaling}
		q_n\simeq
		\frac{\langle g|D_{\mathrm{XUV}}|\mathcal R_n\rangle}
		{\sqrt{\pi\Gamma_{nn}/2}\,
			\langle g|D_{\mathrm{XUV}}|X,\varepsilon_\alpha\rangle}
		\bigg|_{\varepsilon_\alpha=E_{\mathcal R_n}-E_X}.
	\end{equation}
	Because the relative values of \(\langle g|D_{\mathrm{XUV}}|\mathcal R_n\rangle\) are taken from the measured line areas and their overall scale is fixed by \(q_{5d}\) below, they already include the correction to the discrete-state matrix element due to indirect excitation via the continuum. This correction corresponds to the principal-value term of Fano's theory~\cite{fano1961} and varies slowly across each resonance (Sec.~\ref{sec:sup_laplace_markov} of the SM).      
	We set \(q_{5d}=2\) and \(\langle g|D_{\mathrm{XUV}}|X,\varepsilon_\alpha\rangle=1\), giving \(\langle g|D_{\mathrm{XUV}}|\mathcal R_{5d}\rangle=2.0196\times10^{-2}\). The resulting \(q_n\) values are 1.3--4.2 for the sharp and 0.4--0.6 for the diffuse series, reflecting the larger autoionization widths of the diffuse states. For the extrapolated states, \(|\langle g|D_{\mathrm{XUV}}|\mathcal R_n\rangle|^2\) and \(\Gamma_{nn}\) both scale as \((n^*)^{-3}\), and \(q_n\) is therefore constant.
	
	\FloatBarrier
	\subsection{Numerical Model and Observables}\label{sec:numerical_model}
	
	We solve Eqs.~\eqref{eq:c_g_dot}--\eqref{eq:c_beta_dot} for Gaussian XUV and IR pulses of finite pulse lengths. Our implementation contains the 19 Rydberg states of the Henning series (4d--12d and 5s--14s) listed in Table~\ref{tab:Rydberg_States}. The XUV pulse has a peak intensity of \(10^{10}\)~W/cm\(^2\) and a 4.56-fs FWHM, corresponding to the measured \(0.4\)-eV bandwidth. It is centered at \(t\simeq5.5\)~fs, and the IR pulse, with a peak intensity of \(1\)~TW/cm\(^2\), at \(t\simeq5.5~\mathrm{fs}+t_d\). The coupled equations are propagated with an adaptive fifth/fourth-order Runge--Kutta method, starting several pulse lengths before the pump-pulse peak intensity is reached, when the pulse intensity is small enough to be neglected. We allow for relative and absolute tolerances of \(10^{-10}\) of the probability amplitudes and propagate up to \(t_f=24.7\)~ps, which exceeds the longest lifetime in Table~\ref{tab:Rydberg_States} by 50~\%, where all yields are evaluated. The total norm is conserved to better than \(10^{-6}\) throughout the propagation.

	\FloatBarrier
	\subsubsection{Fixed-Delay Propagation and Continuum Formation}
	
\begin{figure}[!htbp]
		\centering
		\includegraphics[width=0.49\textwidth]{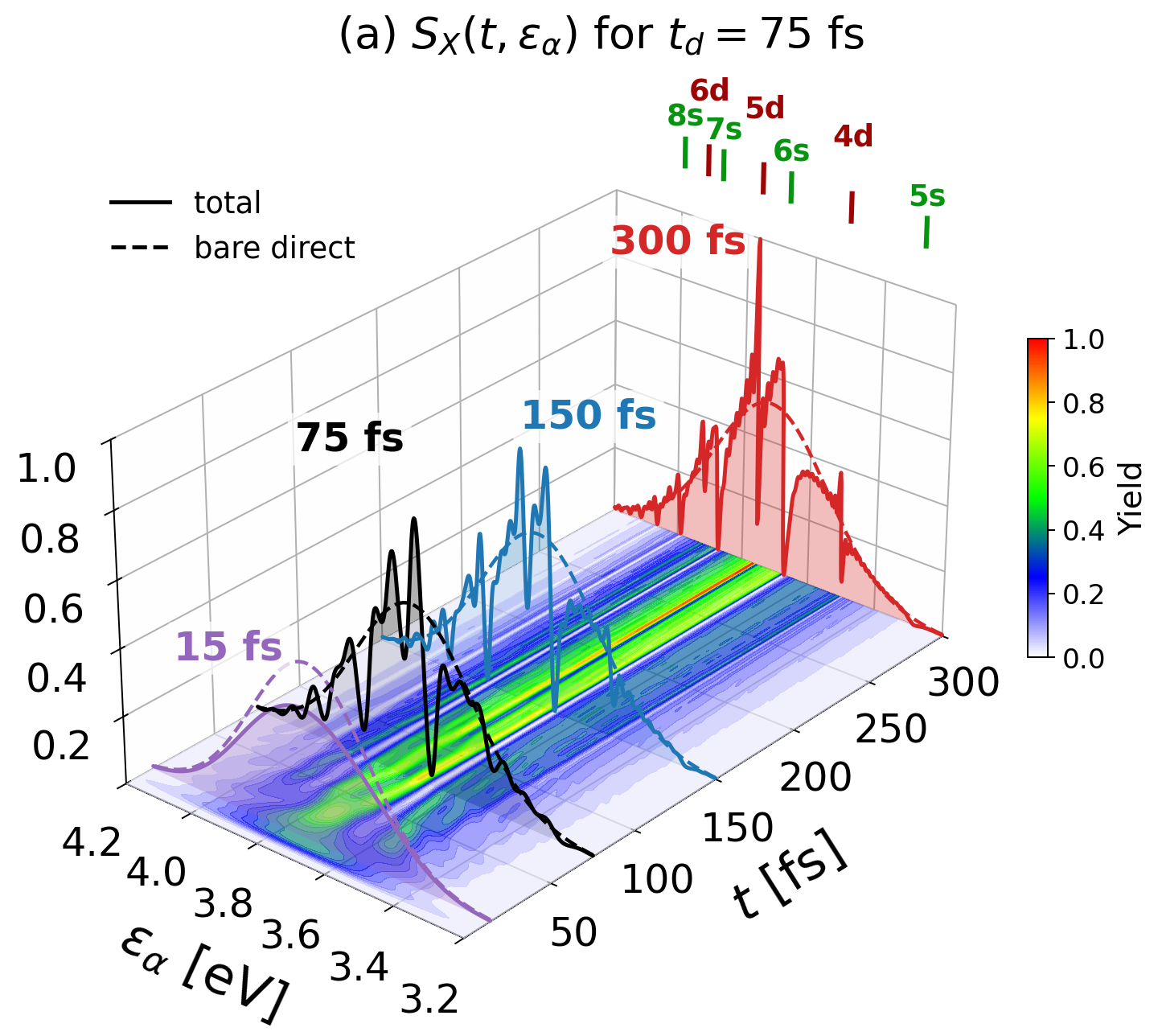}
		\includegraphics[width=0.49\textwidth]{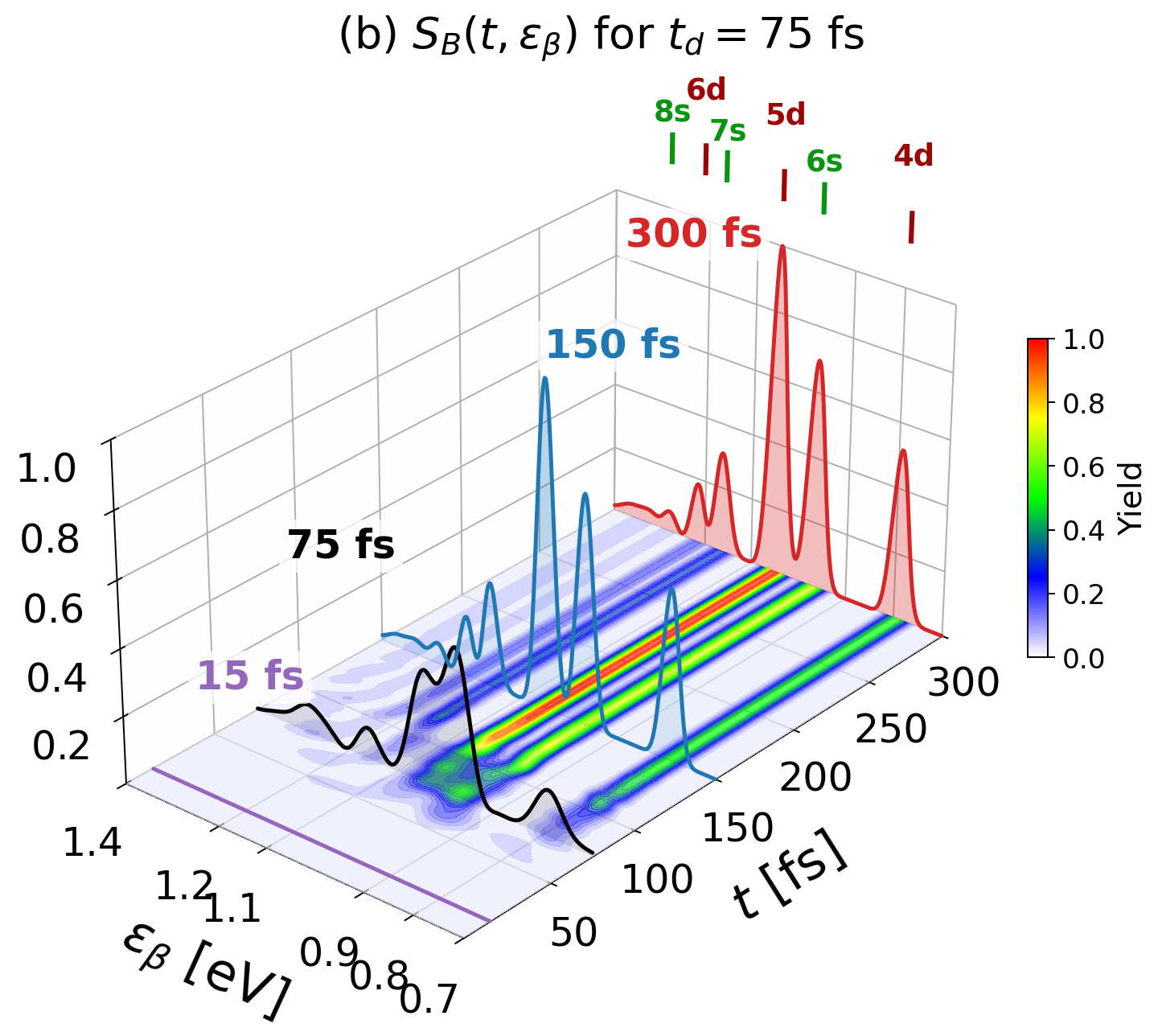}
		\caption{\textbf{Evolution of the photoelectron yield at \(t_d=75\)~fs.} \textbf{(a)} Total \(X\)-channel yield \(S_X(t,\varepsilon_\alpha)\) (solid) and direct-XUV contribution \(|c_X^{\mathrm{dir}}(t,\varepsilon_\alpha)|^2\) (dashed). \textbf{(b)} IR-induced \(B\)-channel yield \(S_B(t,\varepsilon_\beta)\). Floor contours show the evolution from 0 to 300~fs. Line-outs are shown at 15 (purple), 75 (black), 150 (blue), and 300~fs (red). Each panel is normalized to its own maximum. Rear-wall ticks mark the photoelectron energies \(E_{\mathcal R_n}-E_X\) in \textbf{(a)} and \(E_{\mathcal R_n}+\omega_{\mathrm{IR}}-E_B\) in \textbf{(b)} for diffuse (green) and sharp (red) states.}
		\label{fig:single_delay_snapshots}
	\end{figure}

	For a fixed pump--probe delay of \(t_d=75\)~fs,  the formation of the two photoelectron spectra is shown in Fig.~\ref{fig:single_delay_snapshots}. In the \(X\) channel, the same final continuum is reached by direct XUV ionization and by autoionization of the Rydberg wave packet. We therefore combine
	\begin{align}\label{eq:x_amplitude_decomposition}
		c_X(t,\varepsilon_\alpha)
		&=c_X^{\mathrm{dir}}(t,\varepsilon_\alpha)
		+c_X^{\mathrm{auto}}(t,\varepsilon_\alpha), \nonumber\\
		S_X(t,\varepsilon_\alpha)
		&=|c_X^{\mathrm{dir}}|^2+|c_X^{\mathrm{auto}}|^2
		+2\,\mathrm{Re}\!\left[
		(c_X^{\mathrm{dir}})^*c_X^{\mathrm{auto}}\right],
	\end{align}
	where the time- and energy-dependent arguments are suppressed in the second line.

	\begin{figure}[!htbp]
		\centering
		\includegraphics[width=\textwidth]{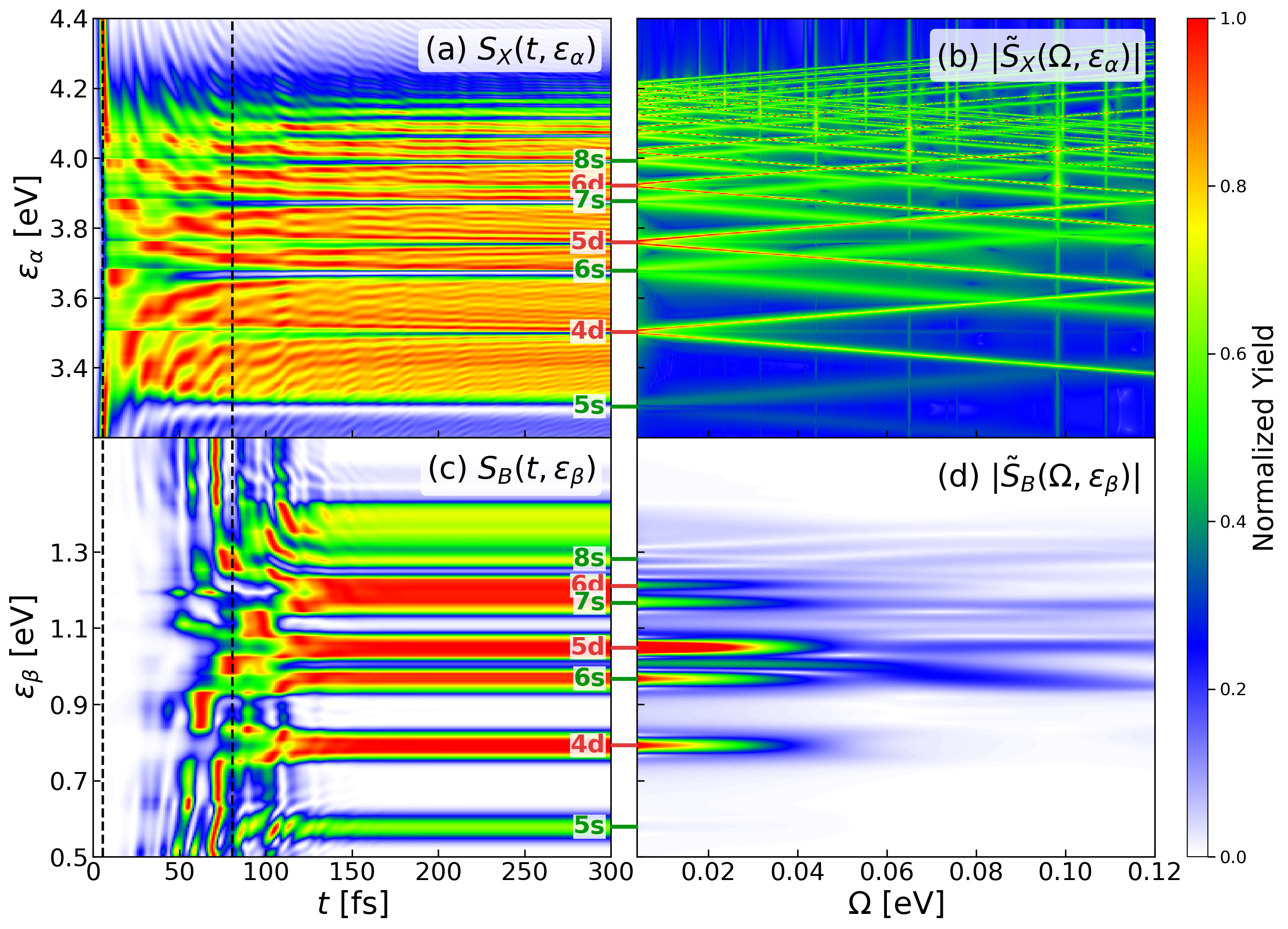}
		\caption{\textbf{Propagation-time evolution of the \(X\)- and \(B\)-channel yields at \(t_d=75\)~fs.} \textbf{(a,c)} Time-domain yields \(S_X(t,\varepsilon_\alpha)\) and \(S_B(t,\varepsilon_\beta)\) during the first 300~fs, with each 1.88-meV-wide photoelectron-energy bin normalized to its maximum over the displayed interval. Vertical dashed lines mark the centers of the XUV pulse at \(t\simeq5.5\)~fs and the IR pulse at \(t\simeq80.5\)~fs.
			\textbf{(b,d)} Magnitudes \(|\tilde S_X(\Omega,\varepsilon_\alpha)|^{0.1}\) and \(|\tilde S_B(\Omega,\varepsilon_\beta)|^{0.9}\) of the corresponding propagation-time Fourier transforms, obtained from the 24.7-ps propagation, after subtracting the DC component and linear trend from each energy row and normalized independently in each panel, with the strongest features saturated. Ticks between the panels mark \(E_{\mathcal R_n}-E_X\) for diffuse (green) and sharp (red) states.}
		\label{fig:xuv_ir_fft_time_sim}
	\end{figure}
	
	After the XUV pulse, the direct contribution is the smooth, nearly Gaussian dashed curve in Fig.~\ref{fig:single_delay_snapshots}(a).
	It no longer changes with propagation time. By 15~fs, a smaller autoionization amplitude has already reached the same continuum. Over much of the spectrum it is out of phase with the direct amplitude, since the interference term in Eq.~\eqref{eq:x_amplitude_decomposition} is predominantly negative. The solid yield (solid line) can therefore lie below the direct yield (dashed lines), even though autoionization provides an additional pathway.
	Between 15 and 300~fs, the dashed background remains unchanged, while the solid curves develop peaks, dips, and asymmetric structures at the marked Rydberg-state energies. These changes directly display the continuing transfer of the Rydberg wave packet into the \(X\) continuum. The broad diffuse-state structures settle first, whereas the narrow sharp-state structures continue to evolve at later times, consistent with the lifetime ordering in Table~\ref{tab:Rydberg_States}.
	Figure~\ref{fig:single_delay_snapshots}(b) contains no signal at 15~fs, because the IR probe pulse has not yet arrived. The final spectrum is created while the probe overlaps the surviving Rydberg wave packet near 75~fs. Once the probe has passed, no further amplitude is transferred to this continuum, such that the 150- and 300-fs line-outs nearly coincide. This stationary \(B\)-channel spectrum contrasts with the continuing evolution of the \(X\)-channel spectrum.

	Panels (a,c) of Fig.~\ref{fig:xuv_ir_fft_time_sim} reproduce the \(X\)- and \(B\)-channel time--energy distributions shown again, differently normalized, as the floor contours in Figs.~\ref{fig:single_delay_snapshots}(a,b). In Fig.~\ref{fig:single_delay_snapshots}, each  panel is normalized to its maximum, whereas in Figs.~\ref{fig:xuv_ir_fft_time_sim}(a,c) each photoelectron-energy row, i.e., each 1.88-meV bin of the continuum energy grid, is divided by its own maximum. This row normalization makes the temporal evolution visible both, near the strongly populated center of the Gaussian excitation spectrum and in its weakly populated wings. Consequently, the colors in panels (a,c) show the evolution at each energy, not the absolute yields at different energies.
	
	All quantum-beat spectra in this work are computed by sampling the electron yield \(Y(\tau,\varepsilon)\) at fixed photoelectron energies \(\varepsilon\) as a function of the time variable \(\tau\) between \(\tau_{\min}\) and \(\tau_{\max}\), 
	\begin{equation}\label{eq:numerical_fft_observable}
		\hat Y(\Omega,\varepsilon)
		=\int_{\tau_{\min}}^{\tau_{\max}} w(\tau)\,[Y(\tau,\varepsilon)-\ell(\tau,\varepsilon)]\,e^{i\Omega\tau}\,d\tau .
	\end{equation}
	At each photoelectron energy, \(\ell(\tau,\varepsilon)=a(\varepsilon)+b(\varepsilon)\,\tau\) is the least-squares linear trend subtracted from that yield. \(w(\tau)\) is a window function, which reduces spectral leakage in the discrete Fourier transform~\cite{Harris1978}. Detrending suppresses the slowly varying contribution near \(\Omega=0\), while the window function reduces artifacts from the finite boundaries of the sampled interval. \(\Omega\) is the frequency conjugate to \(\tau\). For transforms along the propagation time, \(\tau=t\), we use a Hann window and denote the result by a tilde, \(\tilde S\). For transforms along the pump--probe delay, \(\tau=t_d\) (Sec.~\ref{sec:systematic_comparison}), we use a Blackman--Harris window and denote the result by a hat, \(\hat S\).
	
	Figures~\ref{fig:xuv_ir_fft_time_sim}(b,d) show the corresponding Fourier transformations with respect to the propagation time \(t\), \(\tilde S_X(\Omega,\varepsilon_\alpha)\) and \(\tilde S_B(\Omega,\varepsilon_\beta)\),
	as functions of the conjugate frequency \(\Omega\), obtained from Eq.~\eqref{eq:numerical_fft_observable} with \(\tau=t\). To make weak Fourier components visible without changing their positions, the magnitudes are displayed as \(|\tilde S|^\kappa\), with \(\kappa=0.1\) for the \(X\) channel and \(0.9\) for the \(B\) channel.
	The \(X\)-channel V-shaped structures in Fig.~\ref{fig:xuv_ir_fft_time_sim}(b) originate from the equations of motion. After the XUV pulse, the first term of Eq.~\eqref{eq:c_alpha_dot} vanishes, so the direct amplitude \(c_X^{\mathrm{dir}}\) no longer changes [dashed curves in Fig.~\ref{fig:single_delay_snapshots}(a)].
	The second term, \(\sqrt{\Gamma_{nn}/2\pi}\,c_{\mathcal R_n}(t)\,e^{i(E_X+\varepsilon_\alpha-E_{\mathcal R_n})t}\),
	keeps feeding the continuum at energy \(\varepsilon_\alpha\). According to Eq.~\eqref{eq:final_Rydberg_EOM}, \(c_{\mathcal R_n}(t)\) changes only on the time scale of its decay, \(1/\Gamma_{nn}\), apart from the short probe interval.
	The autoionization amplitude that accumulates at \(\varepsilon_\alpha\) therefore rotates in phase relative to the stationary direct amplitude at the continuum--resonance detuning
	\(\Omega=|E_X+\varepsilon_\alpha-E_{\mathcal R_n}|\), and the interference term in Eq.~\eqref{eq:x_amplitude_decomposition} oscillates at this frequency.
	Accordingly, the bright ridges in Fig.~\ref{fig:xuv_ir_fft_time_sim}(b) form V-shaped structures,
	\begin{equation}\label{eq:main_x_v_branch}
		\varepsilon_\alpha=E_{\mathcal R_n}-E_X\pm\Omega .
	\end{equation}
	These V structures are centered at the  on-shell energies \(E_{\mathcal R_n}-E_X\), as indicated for the lowest Henning energies. Opposite arms belonging to states \(n\) and \(m\) intersect at
	\begin{equation}\label{eq:main_x_branch_crossing}
		\Omega_{nm}^{(\times)}
		=\frac{|E_{\mathcal R_m}-E_{\mathcal R_n}|}{2}.
	\end{equation}
	These intersections join the direct--autoionization interference structures
	and are not Rydberg--Rydberg quantum beats. These imprints of interfering autoionization  from different Rydberg states appear as vertical lines at fixed \(\Omega\). Specifically, they arise from the interference between the autoionization amplitudes of different sharp states and lie at their energy spacings (6d--7d, 7d--8d, \ldots). The V-shaped structures and  vertical lines are also present when only the field-free part of the propagation, \(t\geq300\)~fs, is transformed.
	
	The \(S_B\) yield builds up during the probe pulse and remains constant afterwards [Fig.~\ref{fig:single_delay_snapshots}(b)]. The linear detrending removes its DC component, but not the step with which \(S_B\) rises from zero to its final value while the probe pulse, centered at \(t\simeq80.5\)~fs, passes [Fig.~\ref{fig:xuv_ir_fft_time_sim}(c)]. This step is not a numerical artifact: it is the formation of the \(B\)-channel spectrum itself. A straight line fitted over the 24.7-ps window cannot follow it, and its Fourier transform, which decreases as \(1/\Omega\), produces the low-frequency weight in Fig.~\ref{fig:xuv_ir_fft_time_sim}(d).
	Panel (d) thus reflects the formation of the \(B\)-channel spectrum during the probe pulse. It contains neither  oscillations at the Rydberg-state spacings \(|E_{\mathcal R_m}-E_{\mathcal R_n}|\) as a function of the propagation time \(t\) (since \(S_B\) no longer changes after the probe pulse), nor the V-shaped structures of the \(X\) channel. The \(B\)-channel beats appear as a function of the pump--probe delay (Sec.~\ref{sec:systematic_comparison}).

	\FloatBarrier
	\subsubsection{Pump--Probe Imaging of the Autoionization and IR-Ionization Dynamics}

\begin{figure}[!htbp]
		\centering
		\includegraphics[width=\textwidth]{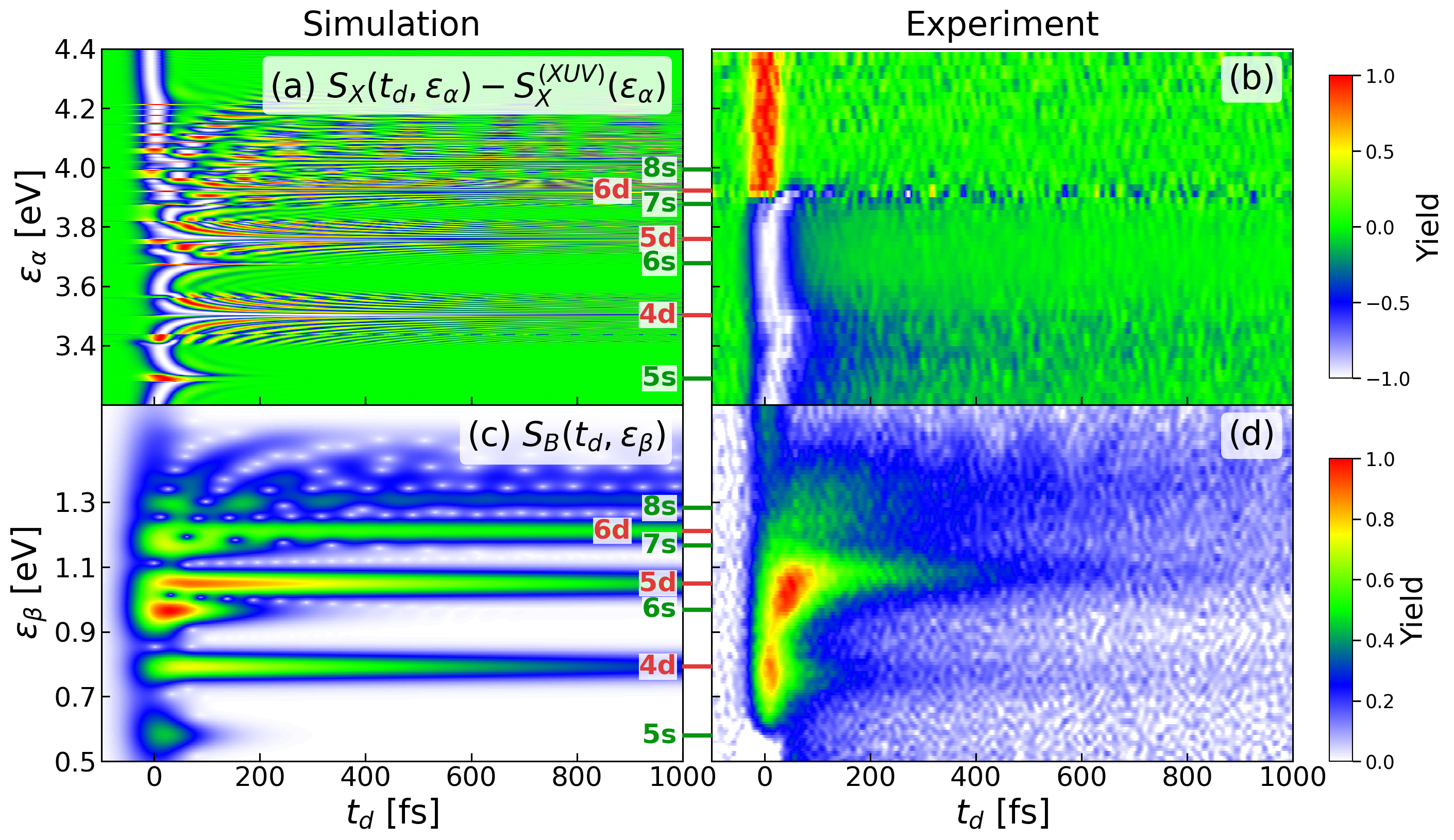}
		\caption{\textbf{Pump--probe-delay-dependent simulated and measured electron yields.}
			\textbf{(a,c)} Simulated and \textbf{(b,d)} measured yields. The measured data were provided by the authors of Ref.~\cite{biswas2024}.
			\textbf{(a,b)} \(X^2\Pi_g\)-channel difference spectrum \(\Delta S_X(t_d,\varepsilon_\alpha)\), normalized at each photoelectron energy to its largest absolute magnitude.
			\textbf{(c,d)} \(B^2\Sigma_u^+\)-channel yield \(S_B(t_d,\varepsilon_\beta)\), displayed as \(S_B^{0.3}\) and \(S_B^{0.8}\), respectively, and normalized independently in each panel. Negative values of the measured difference signal in panel (d) are set to zero.
			Energy ticks between paired panels mark the calculated Henning-series energies; diffuse and sharp states are indicated in green and red, respectively.}
		\label{fig:auto_ir_time_compare}
	\end{figure}

	Figure~\ref{fig:auto_ir_time_compare} compares the calculated and measured delay-dependent electron yields. For the \(X\) channel, we show the difference spectrum
	\begin{equation}\label{eq:delay_difference_spectrum}
		\Delta S_X(t_d,\varepsilon_\alpha)
		=S_X(t_d,\varepsilon_\alpha)-S_X^{(\mathrm{XUV})}(\varepsilon_\alpha),
	\end{equation}
	where \(S_X(t_d,\varepsilon_\alpha)\) is the delay-dependent final yield at the end of the propagation, \(t_f= 24.7\)~ps. We apply the same value for \(t_f\) for all pump--probe delays \(t_d\). \(S_X^{(\mathrm{XUV})}\) is the electron yield produced by the XUV pulse alone. The \(B\)-channel panels show the IR-induced yield \(S_B(t_d,\varepsilon_\beta)\).
	Energy conservation separates the photoelectron-energy windows of the two channels. Direct ionization and autoionization populate the \(X\) continuum at \(\approx3.2\)--\(4.4\)~eV [panels (a,b)], whereas one-photon IR ionization of the Rydberg states by the \(1.59\)-eV probe populates the \(B\) continuum at \(\approx0.5\)--\(1.5\)~eV [panels (c,d)].
	The calculated and the measured \(X\)-channel spectra in Figs.~\ref{fig:auto_ir_time_compare}(a,b) are normalized independently at each photoelectron energy, as in Figs.~\ref{fig:xuv_ir_fft_time_sim}(a,c).
	The calculated and the measured \(B\)-channel spectra in Figs.~\ref{fig:auto_ir_time_compare}(c,d) are normalized to their own maxima and displayed as \(S_B^\kappa\), with \(\kappa=0.3\) for the calculation and \(0.8\) for the measurement for better contrast and comparison.
	
	The XUV-only signal contains the coherent sum of the direct amplitude and the delayed autoionization amplitude [Eq.~\eqref{eq:x_amplitude_decomposition}]. Coupling to the \(B\) continuum [the \(c_B\) term in Eq.~\eqref{eq:final_Rydberg_EOM}] changes the magnitude and phase of the Rydberg contribution before it reaches the \(X\) continuum. The difference spectrum can therefore be positive or negative, producing the alternating regions in Fig.~\ref{fig:auto_ir_time_compare}(a). This is a small correction to a large background. Over the plotted energy window, the calculated maximum of \(|\Delta S_X|\) is \(7.5\%\) of the maximum of \(S_X^{(\mathrm{XUV})}\), and the row normalization enhances this weak structure.
	
	The measured spectra in Fig.~\ref{fig:auto_ir_time_compare}(b) show only a broad change of sign at early delays. They do not reproduce the detailed energy- and delay-dependent structure of the calculation for three reasons. First, the coarser energy sampling and finite energy resolution of the experiment smooth out rapid phase variations. Second, the experiment includes decay channels---notably predissociation and vibrational relaxation---that are absent in our numerical model and modify the relative phase and amplitude of the autoionization contribution. Third, the limited signal-to-noise ratio in the \(X\) channel obscures weaker interference features, because the difference signal is small compared with the XUV-only background. The comparison therefore confirms the energy and delay ranges of the response, while the individual calculated fringes remain a prediction for measurements with higher sensitivity.
	
	The \(B\) channel has no  direct-XUV continuum background. It is populated only when the IR pulse projects the surviving Rydberg wave packet onto the continuum. The short-lived low-\(n\) diffuse Henning states (lifetimes \(\tau_n\le94\)~fs) have largely decayed by \(t_d\simeq200\)~fs, whereas sharp-state contributions persist to later delays. Consistently, the experimental Henning diffuse lines disappear beyond approximately 200~fs, while the visible imprint of the measured 4d, 5d, and 6d Henning sharp lines persists, in agreement with simulation. In the measured map, however, states converging toward the series limit are not individually resolved.
	
	Rydberg signatures are more prominent in the measured \(B\) channel yield [Fig.~\ref{fig:auto_ir_time_compare}(d)] than in the differential \(X\) channel [Fig.~\ref{fig:auto_ir_time_compare}(b)]. The dominant \(5d\) state is particularly informative, as it lies near the center of the XUV spectrum and has one of the largest XUV dipole couplings in Table~\ref{tab:Rydberg_States}.
	In contrast, the corresponding measured \(X\)-channel feature is not discernible although the calculation predicts it. Predissociation of the Henning sharp states, omitted in our model [item (vi) of Sec.~\ref{sec:main_eom}], competes with autoionization. In a competing-decay-rate picture, if state \(n\) has an additional dissociation width \(\Gamma_n^{(\mathrm{diss})}\), only the fraction
	\begin{equation}
		\eta_n=\frac{\Gamma_{nn}}{\Gamma_{nn}+\Gamma_n^{(\mathrm{diss})}}
	\end{equation}
	of its decay reaches the autoionization channel, whereas an early-delay \(B\)-channel signal is reduced only by the survival factor \(e^{-\Gamma_n^{(\mathrm{diss})}t_d}\).
	Biswas \emph{et al.}~\cite{biswas2024} extracted the autoionization and predissociation rates of the \(4d\), \(5d\), and \(6d\) states from the delay dependence of the measured \(B\)-channel yields, resulting in \(\eta_n\simeq0.86\)--\(0.88\). Neglecting predissociation therefore overestimates the sharp-state \(X\)-channel yield by only about 15\%, nearly equally for the three states, and cannot, by itself, explain the missing \(5d\) \(X\)-channel feature in Fig.~\ref{fig:auto_ir_time_compare}(b).
	We attribute its absence mainly to the small differential contrast and the limited energy resolution discussed above.

	\FloatBarrier
	\needspace{6\baselineskip}
	\subsubsection{Fourier-Domain Delay-Range Benchmarking}
	\label{sec:systematic_comparison}
	The quantum-beat spectra discussed in Fig.~\ref{fig:xuv_ir_fft_time_sim}(b,d) are Fourier transformations along the propagation time \(t\) at a single delay and not experimentally observable. We now Fourier transform along the pump--probe delay \(t_d\), which is scanned over a finite delay range in pump--probe experiments. The resulting spectra, \(\Delta\hat S_X(\Omega,\varepsilon_\alpha)\) and \(\hat S_B(\Omega,\varepsilon_\beta)\), are obtained from \(\Delta S_X(t_d,\varepsilon_\alpha)\) and \(S_B(t_d,\varepsilon_\beta)\) according to Eq.~\eqref{eq:numerical_fft_observable} for \(\tau=t_d\) with a Blackman--Harris window.  \(\tau_{\min}\) and \(\tau_{\max}\) are the endpoints of the sampled delay range.
	Because the delay-independent reference \(S_X^{(\mathrm{XUV})}\) is absorbed by the linear trend, \(\Delta\hat S_X\) equals the transform of \(S_X\). Throughout this subsection, \(\Omega\) is conjugate to the pump--probe delay \(t_d\). It is distinct from the propagation-time frequency used in Fig.~\ref{fig:xuv_ir_fft_time_sim}. Each panel has its own display scale and power transformation.
	
\begin{figure}[!hbp]
		\centering
		\includegraphics[width=\textwidth]{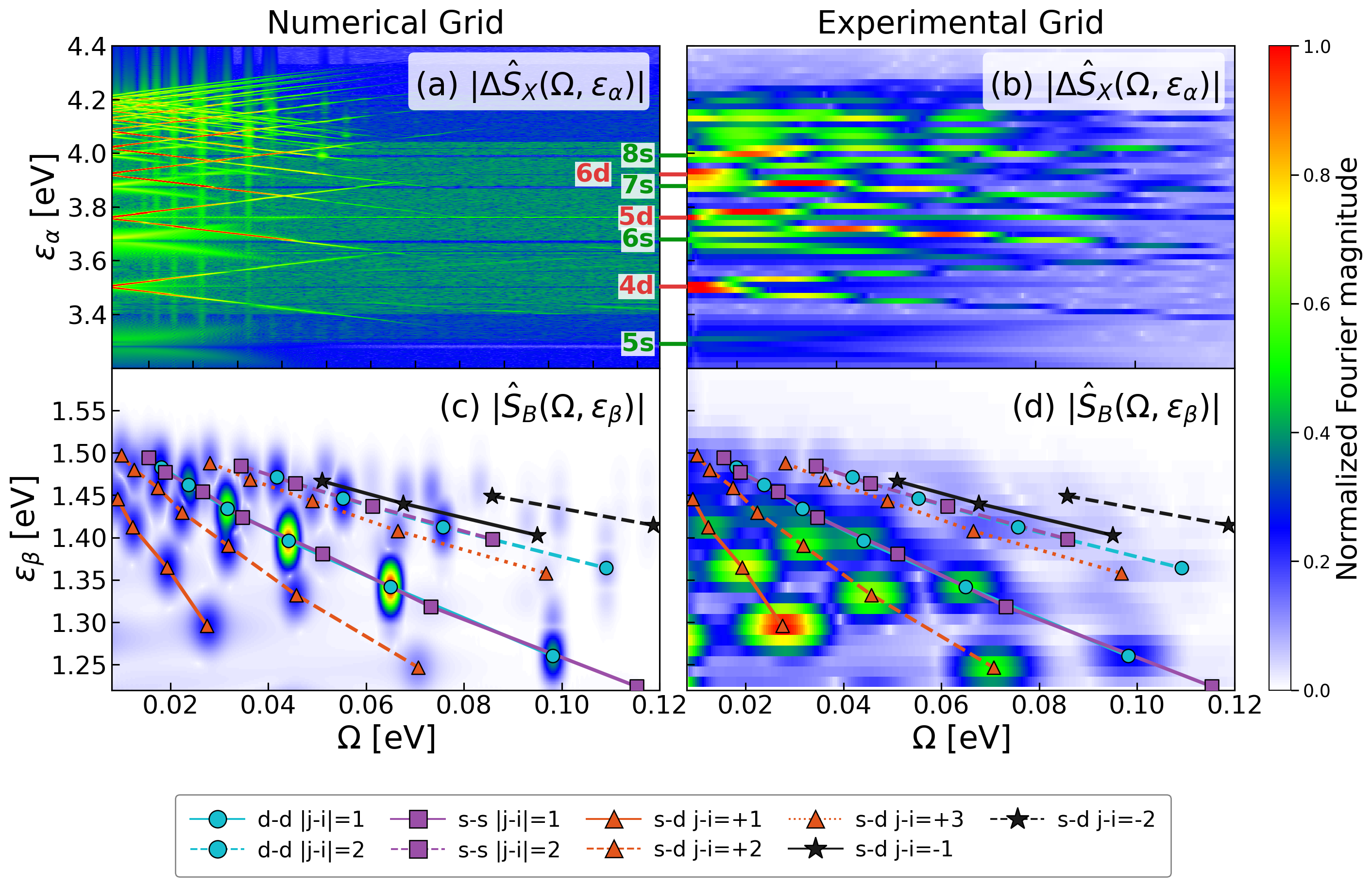}
		\caption{\textbf{Effect of delay sampling on the simulated Fourier spectra.}
			Results for delay ranges of \textbf{(a,c)} \(4.2\)~ps and \textbf{(b,d)} \(1.1\)~ps. The experimental data reported in the present work were accumulated for a delay range of \(1.1\)~ps.
			\textbf{(a,b)} \(X^2\Pi_g\)-channel Fourier magnitudes, displayed as \(|\Delta\hat S_X|^{0.1}\) and \(|\Delta\hat S_X|^{0.2}\), respectively. \textbf{(c,d)} \(B^2\Sigma_u^+\)-channel Fourier magnitudes are plotted  as \(|\hat S_B|^{0.5}\).
			The narrow red diagonal lines in \textbf{(a)} are not added guides, but emerge in the calculated spectra.
			Markers in \textbf{(c,d)} are assignment aids, placed at beat frequencies and pair energies according to Eqs.~\eqref{eq:main_beat_frequency} and \eqref{eq:main_pair_energy}, respectively.
			In \textbf{(c,d)}, cyan circles, purple squares, orange triangles, and black stars denote quantum beats between d--d (sharp--sharp), s--s (diffuse--diffuse), positive--\(n_s-n_d\) s--d, and negative--\(n_s-n_d\) s--d (mixed) Henning Rydberg states, respectively,
			where \(n_s\) and \(n_d\) are the principal quantum numbers of the diffuse and sharp states.
			Solid, dashed, and dotted curves connect pairs with principal-quantum-number separations of 1, 2, and 3. Each panel is normalized independently. Ticks between the panels \textbf{(a)} and \textbf{(b)} refer to Henning series energies, as in Fig.~\ref{fig:xuv_ir_fft_time_sim}.
			The d--d and s--s branches nearly coincide, because at a given binding energy the level spacings of the two Henning series are nearly equal. Where their lines overlap, pairs are distinguished by markers.}	
		\label{fig:auto_ir_fft_compare}
	\end{figure}

	Figure~\ref{fig:auto_ir_fft_compare} applies this procedure to the same calculated yields on two sampling grids. Panels (a,c) use a \(4.2\)-ps delay range with \(2\)-fs delay steps and \(1.88\)-meV photoelectron-energy steps. Panels (b,d) use the \(1.1\)-ps delay range and \(2\)-fs steps of the measured data provided by the authors of Ref.~\cite{biswas2024}, with approximately \(21\)-meV energy bins in the \(X\) window and \(13\)-meV bins in the \(B\) window. The Blackman--Harris-windowed frequency resolution is approximately \(1.9\)~meV for the longer delay range and \(7.1\)~meV for the experimental delay range. The right column is therefore a prediction of how the same calculated response appears under experimental sampling, not measured data.

	\begin{center}\emph{The \(X\) channel: XUV-emission and autoionization}\end{center}
	
	The quantum-beat spectrum in Fig.~\ref{fig:xuv_ir_fft_time_sim}(b) and the delay-Fourier map in Fig.~\ref{fig:auto_ir_fft_compare}(a) show V-shaped \(X\)-channel structures centered at the sharp- and diffuse-state energies. On the dense grid, Fig.~\ref{fig:auto_ir_fft_compare}(a) resolves the V-shaped branches given by Eq.~\eqref{eq:main_x_v_branch} that intersect at \(\Omega_{nm}^{(\times)}\) according to Eq.~\eqref{eq:main_x_branch_crossing}.
	By continuously perturbing the surviving Rydberg population, the IR probe makes the ongoing formation of the \(X\)-channel spectrum visible in the delay-dependent signal. In both spectra, the sharp-state structures are narrow and the diffuse-state structures broad, reflecting their smaller and larger autoionization widths \(\Gamma_{nn}\), respectively (Table~\ref{tab:Rydberg_States}). On the 1.1-ps experimental delay range of Fig.~\ref{fig:auto_ir_fft_compare}(b), the \(7.1\)-meV windowed frequency resolution and approximately \(21\)-meV photoelectron-energy bins broaden and overlap the V arms, leaving mainly horizontal bands at the resonance energies. For the delay range applied in the experiment~\cite{biswas2024}, the V pattern is therefore unlikely to be observable and remains a prediction. Resolving the V-shaped structures requires a longer scan, finer energy sampling, and a sufficient signal-to-noise ratio in the weak differential \(X\) signal.

	\begin{center}\emph{Rydberg--Rydberg beating in the \(X\) channel}\end{center}

	The IR probe also modifies the Rydberg wave packet itself. It transfers population into the \(B\) continuum and, during the probe pulse, coherently returns part of this population [the \(c_B\) term in Eq.~\eqref{eq:final_Rydberg_EOM}]. This return acts as an effective two-photon Raman coupling between  same-parity Henning states. Both processes change the relative amplitudes and phases of the Rydberg states, so that their subsequent autoionization oscillates with delay at the Rydberg-state spacings \(|E_{\mathcal R_m}-E_{\mathcal R_n}|\). Because these frequencies do not depend on the photoelectron energy \(\varepsilon_\alpha\), the resulting Rydberg--Rydberg beats appear as faint, nearly vertical lines in Fig.~\ref{fig:auto_ir_fft_compare}(a), rather than as V-shaped branches. They are much weaker than the direct--autoionization structures and not resolved due to the comparatively short experimental delay range in Fig.~\ref{fig:auto_ir_fft_compare}(b). Such beats become the dominant signal in the Raman-interferometry scheme of Ref.~\cite{Plunkett2022Raman}, where a femtosecond short-wave-infrared probe pulse drives the Raman redistribution within the autoionizing wave packet, and where the beats appear as delay-dependent oscillations in the yields of the narrow autoionization lines of the differential electron spectrum. 
	Even on the \(4.2\)-ps delay range in Fig.~\ref{fig:auto_ir_fft_compare}(a), these beats remain faint, for two reasons. First, the Raman redistribution between the Henning states proceeds only through the \(B\) continuum [item (v) of Sec.~\ref{sec:main_eom}] and is weak, whereas in Ref.~\cite{Plunkett2022Raman} it proceeds through a sum over discrete intermediate states as well as an integral over the continuum. Second, the Henning resonances lie in a strong direct-ionization continuum (\(q_n=0.4\)--4.2, Table~\ref{tab:Rydberg_States}), so that the probe-induced change of the autoionization amplitude enters \(\Delta S_X\) mainly through its interference with the unchanged direct amplitude, producing the V-shaped structures in Fig.~\ref{fig:auto_ir_fft_compare}(a). In Ref.~\cite{Plunkett2022Raman}, the wave packet is excited by an EUV pulse combined with a near-infrared pulse. The EUV photon energy lies below the argon ionization threshold, so that direct ionization by the pump is weak. Hence, no direct--autoionization interference and no V-shaped structures arise there, and the Raman-induced Rydberg--Rydberg beats appear as a clean signal. In CO\(_2\), in contrast, the weak Rydberg--Rydberg beats are masked by the much stronger V-shaped direct--autoionization structures. Observing Rydberg--Rydberg beats in the yields of the narrow autoionization lines of the \(X\)-channel spectrum of CO\(_2\), as in Ref.~\cite{Plunkett2022Raman}, would thus require a probe that couples the Henning states through intermediate \(np\) or \(nf\) Rydberg states, long delay scans, and an electron-energy resolution that separates the narrow sharp-state lines from the direct-ionization background.

	\begin{center}\emph{The \(B\) channel: coherent IR ionization from Rydberg states}\end{center}
	
	Figures~\ref{fig:auto_ir_fft_compare}(c) and \ref{fig:auto_ir_fft_compare}(d) show quantum beats due to the ionization of excited CO$_2$ molecules by IR probe pulses from different coherently populated Rydberg levels.
	According to Eq.~\eqref{eq:c_beta_dot}, each Rydberg state feeds the \(B\) continuum only while the probe pulse is on, near \(t\simeq t_d\), and carries the phase factor \(e^{-iE_{\mathcal R_n}t_d}\)  into \(c_B\). In \(S_B=|c_B|^2\), the contributions of two coherently populated states \(|\mathcal R_n\rangle\) and \(|\mathcal R_m\rangle\) that reach the same final energy therefore interfere with the relative phase \((E_{\mathcal R_m}-E_{\mathcal R_n})t_d\). They produce a yield oscillation with beat frequencies
	\begin{equation}\label{eq:main_beat_frequency}
		\Omega_{nm}=|E_{\mathcal R_m}-E_{\mathcal R_n}|,
	\end{equation}
	localized where the energy-conservation windows of both states, set by the bandwidth of the IR probe in Eq.~\eqref{eq:c_beta_dot}, overlap, i.e., near the pair-averaged photoelectron energy
	\begin{equation}\label{eq:main_pair_energy}
		\bar\varepsilon^{(B)}_{nm}
		\simeq\frac{E_{\mathcal R_n}+E_{\mathcal R_m}}{2}
		+\omega_{\mathrm{IR}}-E_B.
	\end{equation}
	Since Rydberg-state amplitudes decay with the approximate rate \(\Gamma_{nn}/2\) [the \(m=n\) term of Eq.~\eqref{eq:final_Rydberg_EOM}], the coherence decays at \(\gamma_{nm}\simeq(\Gamma_{nn}+\Gamma_{mm})/2\). Thus \(\Omega_{nm}\) measures the Rydberg-state splitting, \(\bar\varepsilon^{(B)}_{nm}\) specifies the continuum energy onto which the pair is projected, and \(1/\gamma_{nm}\) sets its intrinsic duration. Quantum-beat assignments are added as overlays in Figs.~\ref{fig:auto_ir_fft_compare}(c) and \ref{fig:auto_ir_fft_compare}(d). As these two panels demonstrate, projections along either the photoelectron energy or the quantum-beat frequency comprise multiple pairs of beating states. The two-dimensional panels are significantly more selective than either projection alone~\cite{Plunkett2022Raman,plunkett2023metrology}. We denote sharp--sharp, diffuse--diffuse, and mixed sharp--diffuse coherences by d--d, s--s, and s--d, respectively. For mixed pairs, the sign of \(n_s-n_d\), where \(n_s\) and \(n_d\) are the principal quantum numbers of the diffuse and sharp members, distinguishes the two branches.

	\begin{center}\emph{Widths and intensity of Rydberg state quantum beats}\end{center}
	
	The long-lived high-\(n\) sharp states produce narrow d--d beats in Figs.~\ref{fig:auto_ir_fft_compare}(c,d). Mixed s--d beats are broader, because diffuse Henning states decay quickly, while s--s beats are generally broad and weak. Each diffuse \(ns\sigma_g\) level lies just below a sharp \((n-1)d\sigma_g\) level, and their separation decreases toward the common series limit~\cite{fidler2022}. Higher-\(n\) mixed pairs (for example, 9d--10s at 0.012~eV and 10d--11s at 0.009~eV) therefore crowd near the low-frequency edge of Figs.~\ref{fig:auto_ir_fft_compare}(c,d), \(\Omega\lesssim0.01\)~eV, where the subtraction of the linear trend suppresses slowly varying signals, whereas lower-\(n\) pairs such as \(7d\)--\(8s\) and \(8d\)--\(9s\) (at 0.028~eV and 0.019~eV) remain resolved on the numerical grid.
	
	For a delay range of \(4.2\)~ps in Fig.~\ref{fig:auto_ir_fft_compare}(c), adjacent d--d beats dominate: their long coherence times allow the signal to accumulate through much of the scan. For the \(1.1\)-ps experimental delay range in Fig.~\ref{fig:auto_ir_fft_compare}(d), the time series ends before these coherences have decayed, whereas the short-lived mixed pairs have already contributed most of their signal. The Blackman--Harris window also changes their relative temporal weight, so mixed pairs become more prominent, with \(7d\)--\(8s\) the strongest signal. Within each family, larger principal-number separations generally give weaker beats because the Gaussian IR probe has less spectral overlap for states with larger energy splittings, such that both states ionize into the same final-electron energy. Principal-number separation is only an approximate guide to the splitting, and excitation strengths, decay rates, and overlapping features also affect the amplitudes. The shorter delay range broadens the beat frequencies, while coarse photoelectron-energy sampling merges nearby mixed-pair coordinates into a band that cannot be assigned to a single coherence. Beat prominence therefore depends on both the molecular dynamics and the sampling window.

	\section{Conclusion}\label{sec:conclusion}
	
	We have developed and tested a general framework that links molecular photoionization dynamics in ultrashort XUV and IR laser pulses
	to complementary observables in pump-probe experiments, the delay-dependent photoelectron yields in the autoionization (\(X\)) and IR-ionization (\(B\)) channels and  quantum beats between coherently excited Rydberg states that they encode.
	By solving the time-dependent Schr\"odinger equation within a close-coupling expansion, we establish a direct mapping of underlying quantum interferences
	within coherent Rydberg wave packets
	to signatures that are observable in pump-probe-delay dependent emitted electron spectra.
	We find that the Rydberg-wave-packet dynamics can be imaged in two complementary ways:
	(i) The autoionization region of the electron spectrum ($X$ channel)  maps the coherent population and decay of autoionization resonances against direct ionization from the molecular ground state by the pump pulse. This happens on the time scale of the Rydberg-state lifetimes.
	Upon Fourier transformation along the propagation time, the electron-emission evolution displays V-shaped branches in the frequency--energy plane as a characteristic signature.
	These structures and the evolving Fano profile encode the continuum--resonance detuning and the autoionization widths.
	(ii) IR-induced photoionization out of Rydberg levels of the neutral molecule in the  $B$ continuum channel images Rydberg--Rydberg coherences at the pair-averaged energy $\bar\varepsilon^{(B)}_{nm}$ and state-specific quantum-beat frequencies.
	The IR pulse projects pairs of coherently populated Rydberg states onto a common continuum. The \(B\) channel therefore reveals Rydberg--Rydberg phase differences, whereas the dominant \(X\)-channel structure encodes direct--autoionization phase differences.
	
	This complementary assessment of the two ionization channels is essential for uniquely interpreting attosecond pump--probe measurements with multiple coherently excited Rydberg states.
	The \(B\) channel alone yields the Rydberg-state splittings \(\Omega_{nm}\) and coherence lifetimes, but not the asymmetry parameters \(q_n\), which follow from the interference of autoionization with direct ionization into the \(X\) continuum. The \(X\) channel alone shows the autoionization widths \(\Gamma_{nn}\) and the interference with direct ionization.
	The two channels also differ in spectral resolution. In the $B$ channel, the probe pulse releases an electron, such that each spectral feature carries the probe-pulse bandwidth---$\simeq37$~meV for a 50-fs pulse---and is tied by Fourier duality to the probe-pulse duration.
	In the $X$ channel, the electron is released by spontaneous autoionization, so that each resonance-associated  feature carries the natural autoionization width $\Gamma_{nn}$ of the originating state (sub-meV for long-lived Henning sharp states).
	The total \(X\)-channel spectrum additionally contains the broad direct-ionization background and its interference with autoionization. The \(X\) channel can therefore display narrow level structures, while the near-background-free \(B\) channel allows the cleaner measurement of weak Rydberg coherences.
	The ability of the \(X\) channel to resolve features as narrow as the natural autoionization widths, however, applies to the Rydberg-state energies and widths, not to Rydberg--Rydberg beats. The IR probe redistributes the Rydberg populations by Raman transitions only through the \(B\) continuum, and the resulting beats are weak compared with the direct--autoionization interference in the \(X\) channel. Resolving them, as in the Raman-interferometry scheme of Ref.~\cite{Plunkett2022Raman}, would require a probe that couples the Rydberg states through bound intermediate states (Sec.~\ref{sec:systematic_comparison}).

	In an application to CO$_2$ we illustrated both strengths and limits of our numerical model. Short-lived Henning diffuse states dominate the emission dynamics at early times, while long-lived sharp states persist over the whole delay range and govern long-time (long pump-probe delay) response.
	The relative prominence of the resulting quantum beats is therefore not an intrinsic state property; it depends jointly on the coherence lifetime, probe bandwidth, delay span, and energy sampling. A delay scan shorter than the longest coherence times suppresses long-lived pairs and emphasizes short-lived ones.
	For testing our numerical predictions, the delay range is the primary limiting factor.
	The $B$ channel is signal-favorable (small background), whereas the $X$-channel structure is differential and much weaker, requiring both longer delay scans and a larger signal-to-noise ratio.
	
	A detailed quantitative comparison with experiment remains beyond the present model.
	The numerical application retains one ionic vibrational state and one effective photoelectron partial wave per channel, and only the dominant \(X\) autoionization channel. Quantitative amplitudes and linewidths will require additional ionic continua, vibrational levels, explicit nuclear wave-packet motion, predissociation, rotational averaging, and symmetry-dependent couplings. Two suggested refined measurements would directly constrain the leading uncertainties:
	(i) extending the state-resolved \(B\)-channel decay measurements of Ref.~\cite{biswas2024} from \(4d\)--\(6d\) to higher-\(n\) sharp states would determine their predissociation rates, and
	(ii) a resolved photoabsorption peak-to-background ratio would determine the Fano parameter \(q_n\).
	The framework established here provides a versatile basis towards such extensions and a general route to interpret time-resolved molecular Rydberg dynamics. By correlating photoelectron kinetic energies with Rydberg-state quantum-beat frequencies, this work establishes a route to imaging pump-excited Rydberg wave packets in molecules, and may prompt future studies of ultrafast processes in complex quantum systems, including photoionization and autoionization as well as coherent control of molecular wave packets.

	\begin{acknowledgments}
		This work was supported by the Chemical Sciences, Geosciences, and Biosciences Division, Office of Basic Energy Sciences, Office of Science, US Department of Energy under Award DE-FG02-86ER13491. U.T. acknowledges partial support through NSF grant PHY 2409183. E.S. acknowledges start-up funds from the College of Science and Mathematics (CSM) and the Office of Research at Kennesaw State University. We thank D. Biswas and A. Sandhu for providing the experimental data of Ref.~\cite{biswas2024}.
	\end{acknowledgments}

	\clearpage
	\bibliography{references.bib}
	
	\pagebreak
	\section*{Supplementary Material}
	\input{Supplementary.tex}

\end{document}

%% file: Supplementary.tex
\newcommand{\beginsupplement}{%
	\setcounter{section}{0}
	\renewcommand{\thesection}{S\arabic{section}}%
	\renewcommand{\theHsection}{S\arabic{section}}%
	\renewcommand{\theHsubsection}{S\arabic{section}.\arabic{subsection}}%
	\renewcommand{\theHsubsubsection}{S\arabic{section}.\arabic{subsection}.\arabic{subsubsection}}%
	\setcounter{table}{0}
	\renewcommand{\thetable}{S\arabic{table}}%
	\renewcommand{\theHtable}{S\arabic{table}}%
	\setcounter{figure}{0}
	\renewcommand{\thefigure}{S\arabic{figure}}%
	\renewcommand{\theHfigure}{S\arabic{figure}}%
	\setcounter{equation}{0}
	\counterwithin*{equation}{section}%
	\renewcommand{\theequation}{S\arabic{section}.\arabic{equation}}%
	\renewcommand{\theHequation}{S\arabic{section}.\arabic{equation}}%
	\hypersetup{pageanchor=false}%
	\setcounter{page}{1}
}

\beginsupplement

\section{Theory}\label{sec:sup_theory}

This Supplementary Material outlines the general theoretical framework used to describe the dynamics of molecular Rydberg states under combined XUV-pump and IR-probe laser fields. The formulation allows an arbitrary number of Rydberg resonances and an arbitrary number of ionization channels. Atomic units are used throughout.

\subsection{Hamiltonian and channel basis}

Let \(K_N\) and \(K_e\) denote the nuclear and electronic kinetic-energy operators, respectively, and let \(V_{NN}\), \(V_{Ne}\), and \(V_{ee}\) denote the nucleus--nucleus, nucleus--electron, and electron--electron interactions. The electronic Hamiltonian is
\begin{equation}
	H_e=K_e+V_{NN}+V_{Ne}+V_{ee}.
\end{equation}
Let \(P_a\) be the orthogonal projector onto electronic channel \(a\), with \(P_aP_b=\delta_{ab}P_a\) and \(\sum_aP_a=I\). We define the channel Hamiltonian \(H_{\mathrm{ch}}\) and the configuration interaction \(V_{\mathrm{CI}}\) by
\begin{align}\label{eq:electronic_Hamiltonian}
	H_{\mathrm{ch}}&=K_N+\sum_aP_aH_eP_a,\nonumber\\
	V_{\mathrm{CI}}&=\sum_{a\neq b}P_aH_eP_b.
\end{align}
Adding the channel-diagonal and channel-off-diagonal blocks gives the full field-free molecular Hamiltonian,
\begin{equation}\label{eq:fieldfree_partition}
	H_{\mathrm{ch}}+V_{\mathrm{CI}}=K_N+H_e.
\end{equation}
Thus, \(H_{\mathrm{ch}}\) contains the nuclear kinetic energy and the channel-diagonal blocks of \(H_e\), whereas \(V_{\mathrm{CI}}\) contains the channel-off-diagonal blocks.

Within the Born--Oppenheimer approximation, the electronic channel functions are obtained at fixed nuclear geometry from
\begin{equation}\label{eq:Electronic_SE}
	\left(\sum_aP_aH_eP_a\right)|\psi_i(\vec R)\rangle
	=U_i(\vec R)|\psi_i(\vec R)\rangle,
\end{equation}
where \(\vec R=(R_1,R_2,\ldots)\) denotes the internuclear coordinates, \(|\psi_i(\vec R)\rangle\) is the \(i\)th adiabatic electronic state, and \(U_i(\vec R)\) is the corresponding potential-energy surface. For each channel-diagonal block, before mixing by \(V_{\mathrm{CI}}\), the molecular channel function is written as a product of electronic and nuclear factors. Substitution into the eigenvalue equation for \(H_{\mathrm{ch}}\), with the nonadiabatic derivative couplings generated by the action of \(K_N\) on the \(\vec R\)-dependent electronic functions neglected (Born--Oppenheimer approximation), gives the channel nuclear Schr\"odinger equation
\begin{equation}\label{eq:nuclear_SE}
	\left[K_N+U_i(\vec R)-U_i(\vec R_i^{(e)})\right]
	|\chi_{i,\bm\nu_i}\rangle
	=\varepsilon^{(\mathrm{vib})}_{i,\bm\nu_i}|\chi_{i,\bm\nu_i}\rangle,
\end{equation}
where \(\bm\nu_i\) collectively denotes the vibrational quantum numbers and \(\varepsilon^{(\mathrm{vib})}_{i,\bm\nu_i}\) is measured from the minimum of the corresponding potential-energy surface. The product states
\begin{equation}\label{eq:basis_states}
	|\Psi_{i,\bm\nu_i}\rangle
	=|\psi_i(\vec R)\rangle|\chi_{i,\bm\nu_i}\rangle
\end{equation}
combine the solutions of Eqs.~\eqref{eq:Electronic_SE} and \eqref{eq:nuclear_SE}. Within the same approximation, each product state is an eigenstate of the channel Hamiltonian:
\begin{align}\label{eq:channel_eigenvalue}
	H_{\mathrm{ch}}|\Psi_{i,\bm\nu_i}\rangle
	&=E_{i,\bm\nu_i}|\Psi_{i,\bm\nu_i}\rangle,\nonumber\\
	E_{i,\bm\nu_i}
	&=U_i(\vec R_i^{(e)})+\varepsilon^{(\mathrm{vib})}_{i,\bm\nu_i}.
\end{align}
The states are chosen orthonormal,
\begin{equation}
	\langle\Psi_{i,\bm\nu_i}|\Psi_{j,\bm\nu_j}\rangle
	=\delta_{ij}\delta_{\bm\nu_i\bm\nu_j}.
\end{equation}
For \(N_{\mathrm{nuc}}\) nuclei, there are \(3N_{\mathrm{nuc}}-6\) vibrational coordinates for a nonlinear molecule and \(3N_{\mathrm{nuc}}-5\) for a linear molecule. Expanding \(U_i(\vec R)\) about the equilibrium geometry,
\begin{align}
	U_i(\vec R)\approx&\,U_i(\vec R_i^{(e)})
	+\sum_k\left(\frac{\partial U_i}{\partial R_k}\right)_{\vec R_i^{(e)}}(R_k-R_{i,k}^{(e)})\nonumber\\
	&+\frac12\sum_{k,k'}\left(\frac{\partial^2U_i}{\partial R_k\partial R_{k'}}\right)_{\vec R_i^{(e)}}
	(R_k-R_{i,k}^{(e)})(R_{k'}-R_{i,k'}^{(e)})+\cdots,
\end{align}
the linear term vanishes at equilibrium, and diagonalization of the quadratic form yields the normal-mode contributions to \(\varepsilon^{(\mathrm{vib})}_{i,\bm\nu_i}\). An arbitrary molecular state can therefore be expanded in the electronic-channel and vibrational product states of Eq.~\eqref{eq:basis_states}. No restriction to a single vibrational state is made in the general theory. Projection onto a particular vibrational eigenstate replaces the coordinate-dependent potential in that channel by the constant eigenenergy \(E_{i,\bm\nu_i}\), without identifying the nuclear wave function with a fixed geometry. Rotational motion and orientational averaging are not included [main text, Sec.~\ref{sec:main_eom}, item (vii)].

The dipole interaction \(D(t)\) with the applied laser fields is written in terms of the field-independent dipole-interaction operators \(D_p\), where \(p=\mathrm{XUV},\mathrm{IR}\) labels the two pulses:
\begin{align}\label{eq:Dipole}
	\vec E_p(t)
	&=\vec E_{p}e^{-(t-t_p)^2/(2\tau_p^2)}\cos(\omega_pt+\phi_p-\omega_pt_p), \nonumber\\
	D(t)
	&=\sum_p\left[D_ph_p(t-t_p)e^{i(\omega_pt+\phi_p-\omega_pt_p)}+\mathrm{h.c.}\right],
	\qquad D_p=\vec\mu\cdot\vec E_{p}, \nonumber\\
	h_p(t-t_p)&=\frac12\exp\!\left[-\frac{(t-t_p)^2}{2\tau_p^2}\right],
	\qquad t_{\mathrm{XUV}}=0,\quad t_{\mathrm{IR}}=t_d.
\end{align}
Thus \(h_p(t-t_p)\) is the half-envelope of the \(e^{i\omega_pt}\) component. The factor \(1/2\) ensures that the two frequency components produce a real field with Gaussian envelope \(e^{-(t-t_p)^2/(2\tau_p^2)}\). The pump envelope is centered at \(t=0\), and the probe envelope at the pump--probe delay \(t=t_d\).

\section{XUV Pump and IR Probe Pulses}\label{sec:sup_pulses}

In the presence of an XUV pump pulse and a time-delayed IR probe pulse, the total molecular wave function can be expanded in the channel basis introduced in Sec.~\ref{sec:sup_theory}. The full time-dependent wave function is expressed as a coherent superposition of discrete and continuum states,
\begin{align}\label{eq:sup_Psi_t}
	|\Psi(t)\rangle
	=&\, c_g(t)\, e^{-iE_g t} |g\rangle
	+ \sum_{n} c_{\mathcal{R}_n}(t)\, e^{-iE_{\mathcal{R}_n} t} |\mathcal{R}_n\rangle \nonumber\\
	&+ \sum_{\mathcal{C},\alpha} \int d\varepsilon_{\alpha}\,
	c_{\mathcal{C}}(t,\varepsilon_{\alpha})\, e^{-i(E_{\mathcal{C}}+\varepsilon_{\alpha})t} |\mathcal{C},\varepsilon_{\alpha}\rangle
	+ \sum_{\mathcal{N},\beta} \int d\varepsilon_{\beta}\,
	c_{\mathcal{N}}(t,\varepsilon_{\beta})\, e^{-i(E_{\mathcal{N}}+\varepsilon_{\beta})t} |\mathcal{N},\varepsilon_{\beta}\rangle ,
\end{align}
\noindent
where
\begin{itemize}
	\item \( |g\rangle = |\Psi_{g,\bm{\nu}_g}\rangle \) denotes the ground state of the neutral molecule, with \(E_g\equiv E_{g,\bm\nu_g}\),
	
	\item \( |\mathcal{R}_n\rangle = |\Psi_{n,\bm{\nu}_n}\rangle \) represents an autoionizing Rydberg state, with \(E_{\mathcal R_n}\equiv E_{n,\bm\nu_n}\), where \(n\) labels the Rydberg electronic state and \(\bm\nu_n\) its vibrational level,
	
	\item \( |\mathcal{C},\varepsilon_{\alpha}\rangle = |\Psi_{c,\bm{\nu}_c}\rangle |\varepsilon_{\alpha}\rangle \) denotes a continuum channel coupled to the Rydberg manifold by configuration interaction, with \(E_{\mathcal C}\equiv E_{c,\bm\nu_c}\), where \(c\) labels the residual-ion electronic state and \(\bm\nu_c\) its vibrational level,
	
	\item \( |\mathcal{N},\varepsilon_{\beta}\rangle = |\Psi_{d,\bm{\nu}_d}\rangle |\varepsilon_{\beta}\rangle \) denotes a continuum channel not coupled to the Rydberg manifold by configuration interaction, with \(E_{\mathcal N}\equiv E_{d,\bm\nu_d}\), where \(d\) labels the residual-ion electronic state and \(\bm\nu_d\) its vibrational level.
\end{itemize}
For brevity, each calligraphic label stands for the pair of electronic and vibrational labels, \(\mathcal R_n\equiv(n,\bm\nu_n)\), \(\mathcal C\equiv(c,\bm\nu_c)\), and \(\mathcal N\equiv(d,\bm\nu_d)\). In the general case, the sums therefore run over both, \(\sum_n\equiv\sum_n\sum_{\bm\nu_n}\), \(\sum_{\mathcal C}\equiv\sum_c\sum_{\bm\nu_c}\), and \(\sum_{\mathcal N}\equiv\sum_d\sum_{\bm\nu_d}\), and the equations below keep the same form for any number of vibrational levels. The indices \(\alpha\) and \(\beta\) label photoelectron partial waves.

\noindent
The continuum electron states \( |\varepsilon_{\alpha}\rangle \) and \( |\varepsilon_{\beta}\rangle \) are energy-normalized scattering states satisfying \( \langle \varepsilon | \varepsilon' \rangle = \delta(\varepsilon - \varepsilon') \). The compact energies \(E_g\), \(E_{\mathcal R_n}\), \(E_{\mathcal C}\), and \(E_{\mathcal N}\) are the corresponding Born--Oppenheimer channel eigenenergies defined above.

Within the asymptotic channel representation, the continuum states factorize into residual-ion and energy-normalized photoelectron factors:
\begin{equation}
	|\mathcal{C},\varepsilon_{\alpha}\rangle = |\mathcal{C}\rangle \otimes |\varepsilon_{\alpha}\rangle,
	\qquad
	|\mathcal{N},\varepsilon_{\beta}\rangle = |\mathcal{N}\rangle \otimes |\varepsilon_{\beta}\rangle,
\end{equation}
with corresponding total energies \( E_{\mathcal{C}} + \varepsilon_{\alpha} \) and \( E_{\mathcal{N}} + \varepsilon_{\beta} \).


\subsection{Equations of Motion}\label{sec:sup_eom}

We study the time evolution of the total wave function \( |\Psi(t)\rangle \) under the influence of the total Hamiltonian
\(
H = H_{\mathrm{ch}}+V_{\mathrm{CI}}+D(t),
\)
where \(H_{\mathrm{ch}}+V_{\mathrm{CI}}\) is the field-free molecular Hamiltonian and \(D(t)\) describes the time-dependent dipole interaction with the external XUV and IR laser fields. The time-dependent Schr\"odinger equation (TDSE) reads
\begin{align}
	i\frac{d}{dt} |\Psi(t)\rangle
	=& [H_{\mathrm{ch}}+V_{\mathrm{CI}}+D(t)] |\Psi(t)\rangle \nonumber\\
	=&\, c_g(t) e^{-iE_g t}[H_{\mathrm{ch}}+V_{\mathrm{CI}}+D(t)]|g\rangle
	+ \sum_{n} c_{\mathcal{R}_n}(t) e^{-iE_{\mathcal{R}_n} t}[H_{\mathrm{ch}}+V_{\mathrm{CI}}+D(t)]|\mathcal{R}_n\rangle \nonumber\\
	&+ \sum_{\mathcal{C},\alpha} \int d\varepsilon_{\alpha}\, c_{\mathcal{C}}(t,\varepsilon_{\alpha}) e^{-i(E_{\mathcal{C}}+\varepsilon_{\alpha})t}[H_{\mathrm{ch}}+V_{\mathrm{CI}}+D(t)]|\mathcal{C},\varepsilon_{\alpha}\rangle \nonumber\\
	&+ \sum_{\mathcal{N},\beta} \int d\varepsilon_{\beta}\, c_{\mathcal{N}}(t,\varepsilon_{\beta}) e^{-i(E_{\mathcal{N}}+\varepsilon_{\beta})t}[H_{\mathrm{ch}}+V_{\mathrm{CI}}+D(t)]|\mathcal{N},\varepsilon_{\beta}\rangle .
\end{align}

Projecting onto the channel basis yields coupled equations for the amplitudes. The rotating-wave approximation (RWA) is applied to the XUV couplings \(g\to\mathcal R_n\) and \(g\to\mathcal C\) and to the IR bound--continuum couplings \(\mathcal R_n\to\mathcal N\). For the latter, the counter-rotating component would lead to \(\varepsilon_\beta=E_{\mathcal R_n}-\omega_{\mathrm{IR}}-E_{\mathcal N}<0\), which is energetically closed. Both frequency components are retained in the IR bound--bound couplings \(\mathcal R_m\leftrightarrow\mathcal R_n\), which are set to zero for CO\(_2\) [item (v) of the CO\(_2\) specialization in the main text]. IR couplings between \(\mathcal C\) and \(\mathcal R_n\) and IR continuum--continuum transitions are not included. We choose time-reversal-symmetric channel phases such that the bare dipole and configuration-interaction matrix elements used in the model are real. Each channel function in Eq.~\eqref{eq:sup_Psi_t} is defined only up to an overall phase. We adopt the standard time-reversal-symmetric choice used in Fano theory~\cite{fano1961,FanoCooper1968}.
Disregarding external magnetic fields and spin--orbit coupling, \(H_{\mathrm{ch}}+V_{\mathrm{CI}}\) is real in coordinate representation, so bound-channel functions \(|g\rangle\), \(|\mathcal{R}_n\rangle\), and continuum channels may be chosen as real standing-wave states~\cite{LaneThomas1958}. For linearly polarized fields in the length gauge, \(D_{\mathrm{XUV}}\) and \(D_{\mathrm{IR}}\) are also real once carrier phases and envelopes are factored out [Eq.~\eqref{eq:Dipole}].
Therefore the length-gauge dipole and configuration-interaction matrix elements,
\[
\langle g|D_{\mathrm{XUV}}|\mathcal{C},\varepsilon_{\alpha}\rangle
=\langle\mathcal{C},\varepsilon_{\alpha}|D_{\mathrm{XUV}}|g\rangle,
\qquad
\langle\mathcal{R}_n|V_{\mathrm{CI}}|\mathcal{C},\varepsilon_{\alpha}\rangle
=\langle\mathcal{C},\varepsilon_{\alpha}|V_{\mathrm{CI}}|\mathcal{R}_n\rangle,
\]
energy shifts \(F_{nm}\), widths \(\Gamma_{nm}\), and asymmetry parameters defined in Eq.~\eqref{eq:FanoParms_clean} below are real.
The incoming-wave states that describe a detected photoelectron follow from these standing-wave states by a unitary transformation at fixed energy, which for a single continuum reduces to the scattering-phase factor \(e^{-i\delta(\varepsilon)}\)~\cite{Starace1982}. The angle-integrated, energy-resolved yields considered here are invariant under this transformation.

\subsubsection{Ground-state amplitude}

The coefficient \(c_g(t)\) is the probability amplitude of the neutral molecular ground state \( |g\rangle \), whose population is \(|c_g(t)|^2\). It is primarily depleted by one-photon XUV excitation to bound Rydberg states \( |\mathcal{R}_n\rangle \) and to the continuum \( |\mathcal{C},\varepsilon_\alpha\rangle \),
\begin{align}\label{eq:sup_c_g_dot}
i\dot{c}_{g}(t)
=& \sum_{n} c_{\mathcal{R}_n}(t)\, h_{\mathrm{XUV}}(t) e^{i\phi_{\mathrm{XUV}}}
e^{-i(E_{\mathcal{R}_n}-E_g-\omega_{\mathrm{XUV}})t}
\langle g|D_{\mathrm{XUV}}|\mathcal{R}_n\rangle \nonumber\\
&+ \sum_{\mathcal{C},\alpha}\!\int d\varepsilon_{\alpha}\,
c_{\mathcal{C}}(t,\varepsilon_{\alpha})\, h_{\mathrm{XUV}}(t) e^{i\phi_{\mathrm{XUV}}}
e^{-i(E_{\mathcal{C}}+\varepsilon_{\alpha}-E_g-\omega_{\mathrm{XUV}})t}
\langle g|D_{\mathrm{XUV}}|\mathcal{C},\varepsilon_{\alpha}\rangle .
\end{align}
Here \(D_{\mathrm{XUV}}\) is the dipole operator coupling the ground state to excited electronic configurations through absorption of an XUV photon.

\subsubsection{Rydberg-state amplitude}

The coefficient \(c_{\mathcal{R}_n}(t)\) is the probability amplitude of the \(n^{\text{th}}\) autoionizing Rydberg state \( |\mathcal{R}_n\rangle \), whose population is \(|c_{\mathcal R_n}(t)|^2\). This amplitude is driven by:
\begin{enumerate}
\item XUV excitation from the ground state (\( |g\rangle \to |\mathcal{R}_n\rangle \)),
\item IR-induced dipole coupling among different Rydberg states (\( |\mathcal{R}_m\rangle \leftrightarrow |\mathcal{R}_n\rangle \)),
\item Configuration interaction between the discrete Rydberg state and the coupled continuum \( |\mathcal{C},\varepsilon_{\alpha}\rangle \) (responsible for autoionization),
\item IR-induced ionization from the Rydberg state to the \(\mathcal{N}\) continuum \( |\mathcal{N},\varepsilon_{\beta}\rangle \).
\end{enumerate}
The resulting equation of motion is
\begin{align}\label{eq:c_r_dot}
i\dot{c}_{\mathcal{R}_n}(t)
=&\, c_{g}(t) h_{\mathrm{XUV}}(t) e^{-i\phi_{\mathrm{XUV}}}
e^{i(E_{\mathcal{R}_n}-E_g-\omega_{\mathrm{XUV}})t}
\langle g|D_{\mathrm{XUV}}|\mathcal{R}_n\rangle \nonumber\\
&+ \sum_{m\ne n} c_{\mathcal{R}_m}(t) h_{\mathrm{IR}}(t-t_d) e^{i(\phi_{\mathrm{IR}}-\omega_{\mathrm{IR}}t_d)}
e^{-i(E_{\mathcal{R}_m}-E_{\mathcal{R}_n}-\omega_{\mathrm{IR}})t}
\langle \mathcal{R}_m|D_{\mathrm{IR}}|\mathcal{R}_n\rangle \nonumber\\
&+ \sum_{m\ne n} c_{\mathcal{R}_m}(t) h_{\mathrm{IR}}(t-t_d) e^{-i(\phi_{\mathrm{IR}}-\omega_{\mathrm{IR}}t_d)}
e^{-i(E_{\mathcal{R}_m}-E_{\mathcal{R}_n}+\omega_{\mathrm{IR}})t}
\langle \mathcal{R}_n|D_{\mathrm{IR}}|\mathcal{R}_m\rangle \nonumber\\
&+ \sum_{\mathcal{C},\alpha}\!\int d\varepsilon_{\alpha}\,
c_{\mathcal{C}}(t,\varepsilon_{\alpha})
e^{-i(E_{\mathcal{C}}+\varepsilon_{\alpha}-E_{\mathcal{R}_n})t}
\langle \mathcal{R}_n|V_{\mathrm{CI}}|\mathcal{C},\varepsilon_{\alpha}\rangle \nonumber\\
&+ \sum_{\mathcal{N},\beta}\!\int d\varepsilon_{\beta}\,
c_{\mathcal{N}}(t,\varepsilon_{\beta})
h_{\mathrm{IR}}(t-t_d) e^{i(\phi_{\mathrm{IR}}-\omega_{\mathrm{IR}}t_d)}
e^{-i(E_{\mathcal{N}}+\varepsilon_{\beta}-E_{\mathcal{R}_n}-\omega_{\mathrm{IR}})t}
\langle \mathcal{R}_n|D_{\mathrm{IR}}|\mathcal{N},\varepsilon_{\beta}\rangle .
\end{align}
The last two terms describe, respectively, (i) autoionization via configuration interaction (\(V_{\mathrm{CI}}\)) and (ii) IR-induced ionization into the \(\mathcal{N}\) continuum.

\subsubsection{\texorpdfstring{\(\mathcal{C}\)}{C}-continuum amplitude}

The coefficient \(c_{\mathcal{C}}(t,\varepsilon_{\alpha})\) is the probability-amplitude density of the continuum state \( |\mathcal{C},\varepsilon_{\alpha}\rangle \). \(|c_{\mathcal C}(t,\varepsilon_\alpha)|^2\) is the corresponding energy-differential population. These states are \emph{coupled} to Rydberg states through configuration interaction and contribute to Fano interference and autoionization,
\begin{align}\label{eq:sup_c_alpha_dot}
i\dot{c}_{\mathcal{C}}(t,\varepsilon_{\alpha})
=&\, c_{g}(t) h_{\mathrm{XUV}}(t) e^{-i\phi_{\mathrm{XUV}}}
e^{i(E_{\mathcal{C}}+\varepsilon_{\alpha}-E_g-\omega_{\mathrm{XUV}})t}
\langle g|D_{\mathrm{XUV}}|\mathcal{C},\varepsilon_{\alpha}\rangle \nonumber\\
&+ \sum_{n} c_{\mathcal{R}_n}(t)
e^{i(E_{\mathcal{C}}+\varepsilon_{\alpha}-E_{\mathcal{R}_n})t}
\langle \mathcal{C},\varepsilon_{\alpha}|V_{\mathrm{CI}}|\mathcal{R}_n\rangle .
\end{align}

\subsubsection{\texorpdfstring{\(\mathcal{N}\)}{N}-continuum amplitude}

The coefficient \(c_{\mathcal{N}}(t,\varepsilon_{\beta})\) is the probability-amplitude density of the continuum state \( |\mathcal{N},\varepsilon_{\beta}\rangle \). \(|c_{\mathcal N}(t,\varepsilon_\beta)|^2\) is the corresponding energy-differential population. These states are \emph{not} coupled to Rydberg states by configuration interaction but can be populated by the IR probe through dipole transitions from \( |\mathcal{R}_n\rangle \),
\begin{align}\label{eq:sup_c_beta_dot}
i\dot{c}_{\mathcal{N}}(t,\varepsilon_{\beta})
=&\, \sum_{n} c_{\mathcal{R}_n}(t)
h_{\mathrm{IR}}(t-t_d) e^{-i(\phi_{\mathrm{IR}}-\omega_{\mathrm{IR}}t_d)}
e^{-i(E_{\mathcal{R}_n}-E_{\mathcal{N}}-\varepsilon_{\beta}+\omega_{\mathrm{IR}})t}
\langle \mathcal{R}_n|D_{\mathrm{IR}}|\mathcal{N},\varepsilon_{\beta}\rangle .
\end{align}
The discrete and continuum channel functions satisfy
\[
\langle i|j\rangle = \delta_{ij}, \qquad
\langle \mathcal{C},\varepsilon_{\alpha}|\mathcal{N},\varepsilon_{\beta}\rangle=0, \qquad
\langle i|\mathcal{C},\varepsilon_{\alpha}\rangle = \langle i|\mathcal{N},\varepsilon_{\beta}\rangle = 0,
\]
where \( i,j \) label discrete bound states, \( \mathcal{C},\mathcal{N} \) denote ionic channels, and \(\alpha,\beta\) label photoelectron partial waves.
Configuration interaction is included only through matrix elements of the form \( \langle \mathcal{R}_n|V_{\mathrm{CI}}|\mathcal{C},\varepsilon_{\alpha}\rangle \), responsible for autoionization of the Rydberg states, while all other couplings, such as \( \langle g|V_{\mathrm{CI}}|\mathcal{C},\varepsilon_{\alpha}\rangle \) or continuum--continuum interactions, are neglected,
\[
\langle \mathcal{R}_n|V_{\mathrm{CI}}|\mathcal{N},\varepsilon_{\beta}\rangle =
\langle g|V_{\mathrm{CI}}|\mathcal{C},\varepsilon_{\alpha}\rangle =
\langle g|V_{\mathrm{CI}}|\mathcal{N},\varepsilon_{\beta}\rangle = 0.
\]
All relevant coupling pathways are illustrated schematically in Fig.~\ref{fig:Fig1} of the main text.

\subsection{Laplace--Transform Solution and Markov--Fano Reduction}\label{sec:sup_laplace_markov}

We solve the coupled equations using the Laplace transform. The propagation starts at a time \(t_0\) before both pulses, when the molecule is still in its ground state. For the one-sided transform, we measure time from \(t_0\); this shift only moves the pulse centers and adds constant carrier phases, so the equations obtained below do not depend on \(t_0\). In the shifted time, the pulses are centered at \(\bar t_{\mathrm{XUV}}=-t_0\) and \(\bar t_{\mathrm{IR}}=t_d-t_0\), each carrier phase \(\phi_p\) becomes \(\phi_p-\omega_p\bar t_p\), and the constant phases \(e^{-iE_jt_0}\) are absorbed into the amplitudes \(c_j\). At the new origin \(t=0\), the system is in the neutral ground state,
\[
c_g(0)=1,\qquad
c_{\mathcal{R}_n}(0)=0,\qquad
c_{\mathcal{C}}(0,\varepsilon_{\alpha})=0,\qquad
c_{\mathcal{N}}(0,\varepsilon_{\beta})=0.
\]
For any amplitude \(c_j(t)\), where for continuum amplitudes the index \(j\) includes the photoelectron energy, \(j=(\mathcal C,\varepsilon_\alpha)\) or \(j=(\mathcal N,\varepsilon_\beta)\), and any function \(f(t)\), we use
\begin{align}\label{eq:Laplace}
\mathcal{L}\{c_j(t)\} &= \tilde c_j(s) = \int_{0}^{\infty} c_j(t)\,e^{-st}\,dt, \nonumber\\
\mathcal{L}\{\dot{c}_j(t)\} &= s\,\tilde c_j(s) - c_j(0), \nonumber\\
\mathcal{L}\{f(t)\,e^{\pm i\omega t}\} &= \tilde f(s \mp i\omega).
\end{align}
To keep the pump/probe envelopes traceable, we define the \emph{windowed amplitudes}
\[
c^{(\mathrm{XUV})}_j(t)=c_j(t)\,h_{\mathrm{XUV}}(t-\bar t_{\mathrm{XUV}}),\qquad
c^{(\mathrm{IR})}_j(t)=c_j(t)\,h_{\mathrm{IR}}(t-\bar t_{\mathrm{IR}}),
\]
with Laplace transforms \(\tilde c^{(\mathrm{XUV})}_j(s)\) and \(\tilde c^{(\mathrm{IR})}_j(s)\).

\subsubsection{Transformed equations}

Applying Eq.~\eqref{eq:Laplace} to the equations of motion gives the following transformed amplitudes.
\begin{itemize}
\item For the ground state \(\tilde c_g(s)\):
\begin{align}\label{eq:cs_g}
	i\,s\,\tilde c_g(s) - i\,c_g(0)
	&= \sum_n \tilde c^{(\mathrm{XUV})}_{\mathcal{R}_n}\!\big(s+i(E_{\mathcal{R}_n}-E_g-\omega_{\mathrm{XUV}})\big)\,e^{i(\phi_{\mathrm{XUV}}-\omega_{\mathrm{XUV}}\bar t_{\mathrm{XUV}})}\,\langle g|D_{\mathrm{XUV}}|\mathcal{R}_n\rangle \nonumber\\
	&\quad + \sum_{\mathcal{C},\alpha}\!\int d\varepsilon_{\alpha}\,
	\tilde c^{(\mathrm{XUV})}_{\mathcal{C},\varepsilon_{\alpha}}\!\big(s+i(E_{\mathcal{C}}+\varepsilon_{\alpha}-E_g-\omega_{\mathrm{XUV}})\big)\,e^{i(\phi_{\mathrm{XUV}}-\omega_{\mathrm{XUV}}\bar t_{\mathrm{XUV}})}\,\langle g|D_{\mathrm{XUV}}|\mathcal{C},\varepsilon_{\alpha}\rangle .
\end{align}
\item For the Rydberg state \(\tilde c_{\mathcal{R}_n}(s)\):
\begin{align}\label{eq:cs_a}
	i\,s\,\tilde c_{\mathcal{R}_n}(s)
	&= \sum_{\mathcal{N},\beta}\!\int d\varepsilon_{\beta}\,
	\tilde c^{(\mathrm{IR})}_{\mathcal{N},\varepsilon_{\beta}}\!\big(s+i(E_{\mathcal{N}}+\varepsilon_{\beta}-E_{\mathcal{R}_n}-\omega_{\mathrm{IR}})\big)\,e^{i(\phi_{\mathrm{IR}}-\omega_{\mathrm{IR}}\bar t_{\mathrm{IR}})}\,
	\langle \mathcal{R}_n|D_{\mathrm{IR}}|\mathcal{N},\varepsilon_{\beta}\rangle \nonumber\\
	&\quad + \sum_{m\ne n} \tilde c^{(\mathrm{IR})}_{\mathcal{R}_m}\!\big(s+i(E_{\mathcal{R}_m}-E_{\mathcal{R}_n}-\omega_{\mathrm{IR}})\big)\,e^{i(\phi_{\mathrm{IR}}-\omega_{\mathrm{IR}}\bar t_{\mathrm{IR}})}\,
	\langle \mathcal{R}_m|D_{\mathrm{IR}}|\mathcal{R}_n\rangle \nonumber\\
	&\quad + \sum_{m\ne n} \tilde c^{(\mathrm{IR})}_{\mathcal{R}_m}\!\big(s+i(E_{\mathcal{R}_m}-E_{\mathcal{R}_n}+\omega_{\mathrm{IR}})\big)\,e^{-i(\phi_{\mathrm{IR}}-\omega_{\mathrm{IR}}\bar t_{\mathrm{IR}})}\,
	\langle \mathcal{R}_n|D_{\mathrm{IR}}|\mathcal{R}_m\rangle \nonumber\\
	&\quad + \tilde c^{(\mathrm{XUV})}_g\!\big(s-i(E_{\mathcal{R}_n}-E_g-\omega_{\mathrm{XUV}})\big)\,e^{-i(\phi_{\mathrm{XUV}}-\omega_{\mathrm{XUV}}\bar t_{\mathrm{XUV}})}\,
	\langle g|D_{\mathrm{XUV}}|\mathcal{R}_n\rangle \nonumber\\
	&\quad + \sum_{\mathcal{C},\alpha}\!\int d\varepsilon_{\alpha}\,
	\tilde c_{\mathcal{C},\varepsilon_{\alpha}}\!\big(s+i(E_{\mathcal{C}}+\varepsilon_{\alpha}-E_{\mathcal{R}_n})\big)\,
	\langle \mathcal{R}_n|V_{\mathrm{CI}}|\mathcal{C},\varepsilon_{\alpha}\rangle .
\end{align}
\item For the coupled continuum \(\tilde c_{\mathcal{C},\varepsilon_{\alpha}}(s)\):
\begin{align}\label{eq:cs_alpha}
	i\,s\,\tilde c_{\mathcal{C},\varepsilon_{\alpha}}(s)
	&= \tilde c^{(\mathrm{XUV})}_g\!\big(s-i(E_{\mathcal{C}}+\varepsilon_{\alpha}-E_g-\omega_{\mathrm{XUV}})\big)\,e^{-i(\phi_{\mathrm{XUV}}-\omega_{\mathrm{XUV}}\bar t_{\mathrm{XUV}})}\,
	\langle g|D_{\mathrm{XUV}}|\mathcal{C},\varepsilon_{\alpha}\rangle \nonumber\\
	&\quad + \sum_n \tilde c_{\mathcal{R}_n}\!\big(s-i(E_{\mathcal{C}}+\varepsilon_{\alpha}-E_{\mathcal{R}_n})\big)\,
	\langle \mathcal{C},\varepsilon_{\alpha}|V_{\mathrm{CI}}|\mathcal{R}_n\rangle .
\end{align}
\end{itemize}

Eliminating \(\tilde c_{\mathcal{C},\varepsilon_{\alpha}}(s)\) from Eqs.~\eqref{eq:cs_a}--\eqref{eq:cs_alpha} yields
\begin{align}\label{eq:cs_r}
i\,s\,\tilde c_{\mathcal{R}_n}(s)
&= \sum_{\mathcal{N},\beta}\!\int d\varepsilon_{\beta}\,
\tilde c^{(\mathrm{IR})}_{\mathcal{N},\varepsilon_{\beta}}\!\big(s+i(E_{\mathcal{N}}+\varepsilon_{\beta}-E_{\mathcal{R}_n}-\omega_{\mathrm{IR}})\big)\,e^{i(\phi_{\mathrm{IR}}-\omega_{\mathrm{IR}}\bar t_{\mathrm{IR}})}\,
\langle \mathcal{R}_n|D_{\mathrm{IR}}|\mathcal{N},\varepsilon_{\beta}\rangle \nonumber\\
&\quad + \sum_{m\ne n} \tilde c^{(\mathrm{IR})}_{\mathcal{R}_m}\!\big(s+i(E_{\mathcal{R}_m}-E_{\mathcal{R}_n}-\omega_{\mathrm{IR}})\big)\,e^{i(\phi_{\mathrm{IR}}-\omega_{\mathrm{IR}}\bar t_{\mathrm{IR}})}\,
\langle \mathcal{R}_m|D_{\mathrm{IR}}|\mathcal{R}_n\rangle \nonumber\\
&\quad + \sum_{m\ne n} \tilde c^{(\mathrm{IR})}_{\mathcal{R}_m}\!\big(s+i(E_{\mathcal{R}_m}-E_{\mathcal{R}_n}+\omega_{\mathrm{IR}})\big)\,e^{-i(\phi_{\mathrm{IR}}-\omega_{\mathrm{IR}}\bar t_{\mathrm{IR}})}\,
\langle \mathcal{R}_n|D_{\mathrm{IR}}|\mathcal{R}_m\rangle \nonumber\\
&\quad + \tilde c^{(\mathrm{XUV})}_g\!\big(s-i(E_{\mathcal{R}_n}-E_g-\omega_{\mathrm{XUV}})\big)\,e^{-i(\phi_{\mathrm{XUV}}-\omega_{\mathrm{XUV}}\bar t_{\mathrm{XUV}})}\,
\langle g|D_{\mathrm{XUV}}|\mathcal{R}_n\rangle \nonumber\\
&\quad + \sum_{m}\sum_{\mathcal{C},\alpha}\!\int d\varepsilon_{\alpha}\,
\frac{\langle \mathcal{C},\varepsilon_{\alpha}|V_{\mathrm{CI}}|\mathcal{R}_m\rangle\,
	\langle \mathcal{R}_n|V_{\mathrm{CI}}|\mathcal{C},\varepsilon_{\alpha}\rangle}
{(E_{\mathcal{R}_n}-E_{\mathcal{C}})-\varepsilon_{\alpha}+i s}\,
\tilde c_{\mathcal{R}_m}\!\big(s-i(E_{\mathcal{R}_n}-E_{\mathcal{R}_m})\big) \nonumber\\
&\quad + \sum_{\mathcal{C},\alpha}\!\int d\varepsilon_{\alpha}\,
\frac{\langle \mathcal{R}_n|V_{\mathrm{CI}}|\mathcal{C},\varepsilon_{\alpha}\rangle\,
	\langle g|D_{\mathrm{XUV}}|\mathcal{C},\varepsilon_{\alpha}\rangle}
{(E_{\mathcal{R}_n}-E_{\mathcal{C}})-\varepsilon_{\alpha}+i s}\,
e^{-i(\phi_{\mathrm{XUV}}-\omega_{\mathrm{XUV}}\bar t_{\mathrm{XUV}})}\,
\tilde c^{(\mathrm{XUV})}_g\!\big(s-i(E_{\mathcal{R}_n}-E_g-\omega_{\mathrm{XUV}})\big).
\end{align}

\subsubsection{Time domain and memory integrals}

Inverting the Laplace transform and returning to the original time variable yields the Rydberg equation with two memory (convolution) integrals,
\begin{align}\label{eq:dcdt_R}
i\dot{c}_{\mathcal{R}_n}(t)=
&\sum_{\mathcal{N},\beta} \int d\varepsilon_{\beta}\, c_{\mathcal{N}}(t,\varepsilon_{\beta}) h_{\mathrm{IR}}(t-t_d) e^{i(\phi_{\mathrm{IR}}-\omega_{\mathrm{IR}}t_d)} e^{-i(E_{\mathcal{N}}+\varepsilon_{\beta}-E_{\mathcal{R}_n}-\omega_{\mathrm{IR}})t} \langle  \mathcal{R}_n|D_{\mathrm{IR}}|\mathcal{N},\varepsilon_{\beta}\rangle \nonumber \\
&+\sum_{m\neq n} c_{\mathcal{R}_m}(t) h_{\mathrm{IR}}(t-t_d) e^{i(\phi_{\mathrm{IR}}-\omega_{\mathrm{IR}}t_d)} e^{-i(E_{\mathcal{R}_m}-E_{\mathcal{R}_n}-\omega_{\mathrm{IR}})t} \langle \mathcal{R}_m|D_{\mathrm{IR}}| \mathcal{R}_n\rangle \nonumber \\
&+\sum_{m\neq n} c_{\mathcal{R}_m}(t) h_{\mathrm{IR}}(t-t_d) e^{-i(\phi_{\mathrm{IR}}-\omega_{\mathrm{IR}}t_d)} e^{-i(E_{\mathcal{R}_m}-E_{\mathcal{R}_n}+\omega_{\mathrm{IR}})t} \langle \mathcal{R}_n|D_{\mathrm{IR}}| \mathcal{R}_m\rangle \nonumber \\
&+ c_{g}(t) h_{\mathrm{XUV}}(t) e^{-i\phi_{\mathrm{XUV}}} e^{i( E_{\mathcal{R}_n}-E_{g}-\omega_{\mathrm{XUV}})t} \langle g|D_{\mathrm{XUV}}|\mathcal{R}_n\rangle
\nonumber \\
&-ie^{-i\phi_{\mathrm{XUV}}} \sum_{\mathcal{C},\alpha}\int d\varepsilon_{\alpha}\langle  \mathcal{R}_n|V_{\mathrm{CI}}|\mathcal{C},\varepsilon_{\alpha}\rangle \langle g|D_{\mathrm{XUV}}|\mathcal{C},\varepsilon_{\alpha}\rangle I_1(t)
\nonumber \\
&-i \sum_{m} \sum_{\mathcal{C},\alpha}\int d\varepsilon_{\alpha}\langle  \mathcal{R}_n|V_{\mathrm{CI}}|\mathcal{C},\varepsilon_{\alpha}\rangle \langle \mathcal{C},\varepsilon_{\alpha}|V_{\mathrm{CI}}|\mathcal{R}_m\rangle \, I_2(t)
\end{align}
where
\begin{align}\label{eq:I1I2_def}
I_1(t) &= \int_0^t\! dt'\; c_g(t-t')\,h_{\mathrm{XUV}}(t-t')\,
e^{i(E_{\mathcal{R}_n}-E_g-\omega_{\mathrm{XUV}})(t-t')}\,
e^{i(E_{\mathcal{R}_n}-E_{\mathcal{C}}-\varepsilon_{\alpha})t'}, \nonumber\\
I_2(t) &= \int_0^t\! dt'\; c_{\mathcal{R}_m}(t-t')\,
e^{i(E_{\mathcal{R}_n}-E_{\mathcal{R}_m})(t-t')}\,
e^{i(E_{\mathcal{R}_n}-E_{\mathcal{C}}-\varepsilon_{\alpha})t'}.
\end{align}

\subsubsection{Markov approximation and Sokhotski--Plemelj identity}

In Markov approximation, the factors \(c_g(t-t')h_{\mathrm{XUV}}(t-t')e^{i(E_{\mathcal R_n}-E_g-\omega_{\mathrm{XUV}})(t-t')}\) and \(c_{\mathcal R_m}(t-t')e^{i(E_{\mathcal R_n}-E_{\mathcal R_m})(t-t')}\) in Eq.~\eqref{eq:I1I2_def} are evaluated at \(t'=0\), and the upper integration limit is extended to infinity. This is justified when the couplings to the continuum vary slowly with energy across the spectral range of these factors (wide-band limit), so that the memory kernel decays much faster than the factors change. The remaining integral is evaluated with the Sokhotski--Plemelj identity~\cite{Plemelj1908,Sakurai2020},

\[
\int_0^{\infty}\! e^{i\Delta t'}\,dt' = \pi\,\delta(\Delta) + i\,\mathcal{P}\!\left(\frac{1}{\Delta}\right),
\]
which gives
\begin{align}\label{eq:I_1_I_2_Approx_clean}
I_1(t) &\approx i\,c_g(t)\,h_{\mathrm{XUV}}(t)\,e^{i(E_{\mathcal{R}_n}-E_g-\omega_{\mathrm{XUV}})t}
\left[\mathcal{P}\frac{1}{E_{\mathcal{R}_n}-E_{\mathcal{C}}-\varepsilon_{\alpha}} - i\pi\,\delta(E_{\mathcal{R}_n}-E_{\mathcal{C}}-\varepsilon_{\alpha})\right], \nonumber\\
I_2(t) &\approx i\,c_{\mathcal{R}_m}(t)\,e^{i(E_{\mathcal{R}_n}-E_{\mathcal{R}_m})t}
\left[\mathcal{P}\frac{1}{E_{\mathcal{R}_n}-E_{\mathcal{C}}-\varepsilon_{\alpha}} - i\pi\,\delta(E_{\mathcal{R}_n}-E_{\mathcal{C}}-\varepsilon_{\alpha})\right],
\end{align}
where \(\mathcal{P}\) denotes the Cauchy principal value and \(\delta(\cdot)\) is the Dirac delta distribution.

\subsubsection{Fano parameterization}

Defining resonance energy shifts \(F_{nm}\), widths \(\Gamma_{nm}\), and the asymmetry parameter \(q_n\) (indices \(n,m\) label Rydberg states),
\begin{align}\label{eq:FanoParms_clean}
F_{nm} &= \sum_{\mathcal{C},\alpha}\,\mathcal{P}\!\int d\varepsilon_{\alpha}\;
\frac{\langle \mathcal{R}_n|V_{\mathrm{CI}}|\mathcal{C},\varepsilon_{\alpha}\rangle \langle \mathcal{C},\varepsilon_{\alpha}|V_{\mathrm{CI}}|\mathcal{R}_m\rangle}
{E_{\mathcal{R}_n}-E_{\mathcal{C}}-\varepsilon_{\alpha}}, \nonumber\\
\Gamma_{nm} &= \sum_{\mathcal{C},\alpha}\,2\pi\,
\langle \mathcal{R}_n|V_{\mathrm{CI}}|\mathcal{C},\varepsilon_{\alpha}\rangle \langle \mathcal{C},\varepsilon_{\alpha}|V_{\mathrm{CI}}|\mathcal{R}_m\rangle
\Big|_{\varepsilon_{\alpha}=E_{\mathcal{R}_n}-E_{\mathcal{C}}}, \nonumber\\
q_n &= \frac{
	\langle g|D_{\mathrm{XUV}}|\mathcal{R}_n\rangle
	+ \sum_{\mathcal{C},\alpha}\,\mathcal{P}\!\int d\varepsilon_{\alpha}\;
	\frac{\langle \mathcal{R}_n|V_{\mathrm{CI}}|\mathcal{C},\varepsilon_{\alpha}\rangle \langle g|D_{\mathrm{XUV}}|\mathcal{C},\varepsilon_{\alpha}\rangle}
	{E_{\mathcal{R}_n}-E_{\mathcal{C}}-\varepsilon_{\alpha}}
}{
	\displaystyle
	\pi\sum_{\mathcal{C},\alpha}
	\langle \mathcal{R}_n|V_{\mathrm{CI}}|\mathcal{C},\varepsilon_{\alpha}\rangle
	\langle g|D_{\mathrm{XUV}}|\mathcal{C},\varepsilon_{\alpha}\rangle
	\Big|_{\varepsilon_{\alpha}=E_{\mathcal{R}_n}-E_{\mathcal{C}}}
},
\end{align}
we follow the common Fano parameterization~\cite{fano1961}. For several physical continua or partial waves with channel-dependent scattering phases, only the combination of continuum states coupled to the resonance interferes with the resonant amplitude, whereas the orthogonal combinations contribute a noninterfering background~\cite{fano1961}. The scalar real \(q_n\) used in the main text refers to this resonance-coupled effective continuum.
The numerator of \(q_n\) contains the optical matrix element of Fano's modified discrete state~\cite{fano1961,FanoCooper1968,Starace1982},
\[
|\widetilde{\mathcal R}_n\rangle=|\mathcal R_n\rangle+\sum_{\mathcal{C},\alpha}\mathcal P\!\int d\varepsilon_\alpha\;
\frac{|\mathcal{C},\varepsilon_\alpha\rangle\langle \mathcal{C},\varepsilon_\alpha|V_{\mathrm{CI}}|\mathcal R_n\rangle}
{E_{\mathcal{R}_n}-E_{\mathcal{C}}-\varepsilon_{\alpha}},
\]
not of the bare bound state. Its principal-value term
is real in our phase convention.
Here, the continuum phases are chosen so that the configuration-interaction matrix elements are real in the single-effective-continuum representation used below. The origin of the factor \((q_n-i)\) can then be displayed directly, and the direct excitation, principal-value contribution, and on-shell delta contribution combine as
\begin{align}
&\langle g|D_{\mathrm{XUV}}|\mathcal R_n\rangle
+\sum_{\mathcal{C},\alpha}\mathcal P\!\int d\varepsilon_\alpha\,
\frac{
	\langle\mathcal R_n|V_{\mathrm{CI}}|\mathcal{C},\varepsilon_\alpha\rangle
	\langle g|D_{\mathrm{XUV}}|\mathcal{C},\varepsilon_\alpha\rangle
}{E_{\mathcal R_n}-E_{\mathcal{C}}-\varepsilon_\alpha}
\nonumber\\
&\quad
-i\pi\sum_{\mathcal{C},\alpha}
\langle\mathcal R_n|V_{\mathrm{CI}}|\mathcal{C},\varepsilon_\alpha\rangle
\langle g|D_{\mathrm{XUV}}|\mathcal{C},\varepsilon_\alpha\rangle
\Big|_{\varepsilon_\alpha=E_{\mathcal R_n}-E_{\mathcal{C}}}
\nonumber\\
&=\pi(q_n-i)\sum_{\mathcal{C},\alpha}
\langle\mathcal R_n|V_{\mathrm{CI}}|\mathcal{C},\varepsilon_\alpha\rangle
\langle g|D_{\mathrm{XUV}}|\mathcal{C},\varepsilon_\alpha\rangle
\Big|_{\varepsilon_\alpha=E_{\mathcal R_n}-E_{\mathcal{C}}}.
\end{align}
Thus the factor \(-i\) originates from the retarded on-shell delta term. With these quantities, we obtain the closed Rydberg equation of motion, with the XUV pump strength \(\rho_n\) of Eq.~\eqref{eq:sup_rho_n},
\begin{align}\label{eq:sup_final_Rydberg_EOM}
i\,\dot c_{\mathcal{R}_n}(t) + i\frac{\Gamma_{nn}}{2}\,c_{\mathcal{R}_n}(t)
&= \rho_n\,c_g(t)\,h_{\mathrm{XUV}}(t)\,e^{-i\phi_{\mathrm{XUV}}}\,e^{i(E_{\mathcal{R}_n}-E_g-\omega_{\mathrm{XUV}})t} \nonumber\\
&\quad + \sum_{\mathcal{N},\beta}\!\int d\varepsilon_{\beta}\,
c_{\mathcal{N}}(t,\varepsilon_{\beta})\,h_{\mathrm{IR}}(t-t_d)\,e^{i(\phi_{\mathrm{IR}}-\omega_{\mathrm{IR}}t_d)}\,
e^{-i(E_{\mathcal{N}}+\varepsilon_{\beta}-E_{\mathcal{R}_n}-\omega_{\mathrm{IR}})t}\,
\langle \mathcal{R}_n|D_{\mathrm{IR}}|\mathcal{N},\varepsilon_{\beta}\rangle \nonumber\\
&\quad + \sum_{m\neq n} c_{\mathcal{R}_m}(t)\,h_{\mathrm{IR}}(t-t_d)\,e^{i(\phi_{\mathrm{IR}}-\omega_{\mathrm{IR}}t_d)}\,
e^{-i(E_{\mathcal{R}_m}-E_{\mathcal{R}_n}-\omega_{\mathrm{IR}})t}\,
\langle \mathcal{R}_m|D_{\mathrm{IR}}|\mathcal{R}_n\rangle \nonumber\\
&\quad + \sum_{m\neq n} c_{\mathcal{R}_m}(t)\,h_{\mathrm{IR}}(t-t_d)\,e^{-i(\phi_{\mathrm{IR}}-\omega_{\mathrm{IR}}t_d)}\,
e^{-i(E_{\mathcal{R}_m}-E_{\mathcal{R}_n}+\omega_{\mathrm{IR}})t}\,
\langle \mathcal{R}_n|D_{\mathrm{IR}}|\mathcal{R}_m\rangle \nonumber\\
&\quad - \sum_{m\neq n} c_{\mathcal{R}_m}(t)\,e^{i(E_{\mathcal{R}_n}-E_{\mathcal{R}_m})t}\,
i\frac{\Gamma_{nm}}{2}.
\end{align}
The diagonal shifts \(F_{nn}\) are absorbed into the resonance energies. The off-diagonal principal-value shifts \(F_{nm}\), \(m\neq n\), are neglected. The cross widths \(\Gamma_{nm}\), \(m\neq n\), are retained. The principal-value term in the numerator of \(q_n\) is handled the same way. Being a smooth function of energy on the scale of \(\Gamma_{nn}\), it acts as a real additive constant across each resonance and is absorbed into \(\langle g|D_{\mathrm{XUV}}|\mathcal R_n\rangle\), resulting in a dressed bound dipole. Equation~\eqref{eq:q_scaling} of the main text is therefore approximate. Since the bound dipoles are fixed empirically
by the integrated peak areas of the measured photoabsorption spectrum~\cite{HuestisBerkowitz2011}, this contribution is already included.
The XUV pump strength is
\begin{equation}\label{eq:sup_rho_n}
\rho_n
=\pi(q_n-i)
\sum_{\mathcal{C},\alpha}
\langle \mathcal{R}_n|V_{\mathrm{CI}}|\mathcal{C},\varepsilon_{\alpha}\rangle
\langle g|D_{\mathrm{XUV}}|\mathcal{C},\varepsilon_{\alpha}\rangle
\Big|_{\varepsilon_{\alpha}=E_{\mathcal{R}_n}-E_{\mathcal{C}}}.
\end{equation}

In Markov (memory-loss) approximation, the exact normalization in Eq.~\eqref{eq:normalization} of the main text is preserved only to the accuracy of the continuum-elimination and weak-field approximations.
Equation~\eqref{eq:sup_final_Rydberg_EOM} shows that (i) diagonal principal-value terms shift the resonance energies, (ii) on-shell delta terms generate the widths \(\Gamma_{nm}\), and (iii) \(\rho_n\) encodes interference between direct XUV ionization and resonant excitation.

%% file: references.bib
@article{biswas2024,
	title={Ultrafast dynamics of the {Rydberg} states of {CO$_2$}: Autoionization and dissociation lifetimes},
	author={Biswas, Dipayan and Wood, J. K. and Shalaby, Islam and Sandhu, Arvinder},
	journal={Phys. Rev. A},
	volume={110},
	number={4},
	pages={043106},
	year={2024},
	doi={10.1103/PhysRevA.110.043106},
	url={https://doi.org/10.1103/PhysRevA.110.043106}
}

@article{kuerpick1997,
	title = {Resonance formation of hydrogenic levels in front of metal surfaces},
	author = {K{\"u}rpick, P. and Thumm, U. and Wille, U.},
	journal = {Phys. Rev. A},
	volume = {56},
	issue = {1},
	pages = {543--554},
	numpages = {0},
	year = {1997},
	month = {Jul},
	publisher = {American Physical Society},
	doi = {10.1103/PhysRevA.56.543},
	url = {https://link.aps.org/doi/10.1103/PhysRevA.56.543}
}

@article{Thumm2008,
	title = {Time-series analysis of vibrational nuclear wave-packet dynamics in $\text{D}_{2}{}^{+}$},
	author = {Thumm, Uwe and Niederhausen, Thomas and Feuerstein, Bernold},
	journal = {Phys. Rev. A},
	volume = {77},
	issue = {6},
	pages = {063401},
	numpages = {12},
	year = {2008},
	month = {Jun},
	publisher = {American Physical Society},
	doi = {10.1103/PhysRevA.77.063401},
	url = {https://link.aps.org/doi/10.1103/PhysRevA.77.063401}
}

@article{fidler2022,
	title={State-selective probing of {CO$_2$} autoionizing inner valence {Rydberg} states with attosecond extreme ultraviolet four-wave-mixing spectroscopy},
	author={Fidler, Ashley P and Lin, Yen-Cheng and Gaynor, James D and McCurdy, C William and Leone, Stephen R and Lucchese, Robert R and Neumark, Daniel M},
	journal={Phys. Rev. A},
	volume={106},
	number={6},
	pages={063525},
	year={2022},
	publisher={APS},
	doi={10.1103/physreva.106.063525},
	url={https://doi.org/10.1103/physreva.106.063525}
}

@article{Plunkett2022Raman,
	author={Plunkett, Alexander and Alarc{\'o}n, Miguel A. and Wood, J. K. and Greene, Chris H. and Sandhu, Arvinder},
	title={{Raman} Interferometry between Autoionizing States to Probe Ultrafast Wave-Packet Dynamics with High Spectral Resolution},
	journal={Phys. Rev. Lett.},
	volume={128},
	number={8},
	pages={083001},
	year={2022},
	doi={10.1103/physrevlett.128.083001},
	url={https://doi.org/10.1103/physrevlett.128.083001}
}

@article{plunkett2023metrology,
	author={Plunkett, Alexander and Wood, J. K. and Alarc{\'o}n, Miguel A. and Biswas, Dipayan and Greene, Chris H. and Sandhu, Arvinder},
	title={High Resolution Metrology of Autoionizing States through {Raman} Interferences},
	journal={J. Phys.: Conf. Ser.},
	volume={2494},
	pages={012003},
	year={2023},
	doi={10.1088/1742-6596/2494/1/012003},
	url={https://doi.org/10.1088/1742-6596/2494/1/012003}
}

@article{plunkett2019,
	title={Ultrafast {Rydberg}-state dissociation in oxygen: Identifying the role of multielectron excitations},
	author={Plunkett, Alexander and Harkema, Nathan and Lucchese, Robert R and McCurdy, C William and Sandhu, Arvinder},
	journal={Phys. Rev. A},
	volume={99},
	number={6},
	pages={063403},
	year={2019},
	publisher={APS},
	doi={10.1103/physreva.99.063403},
	url={https://doi.org/10.1103/physreva.99.063403}
}

@article{liao2017,
	title={Probing autoionizing states of molecular oxygen with {XUV} transient absorption: Electronic-symmetry-dependent line shapes and laser-induced modifications},
	author={Liao, Chen-Ting and Li, Xuan and Haxton, Daniel J and Rescigno, Thomas N and Lucchese, Robert R and McCurdy, C William and Sandhu, Arvinder},
	journal={Phys. Rev. A},
	volume={95},
	number={4},
	pages={043427},
	year={2017},
	publisher={APS},
	doi={10.1103/physreva.95.043427},
	url={https://doi.org/10.1103/physreva.95.043427}
}

@article{desrier2018,
	title={Ionization dynamics through a {Fano} resonance: Time-domain interpretation of spectral amplitudes},
	author={Desrier, Antoine and Maquet, Alfred and Ta{\"\i}eb, Richard and Caillat, J{\'e}r{\'e}mie},
	journal={Phys. Rev. A},
	volume={98},
	number={5},
	pages={053406},
	year={2018},
	publisher={APS},
	doi={10.1103/physreva.98.053406},
	url={https://doi.org/10.1103/physreva.98.053406}
}

@article{argenti2013,
	title={Photoionization of helium by attosecond pulses: Extraction of spectra from correlated wave functions},
	author={Argenti, Luca and Pazourek, Renate and Feist, Johannes and Nagele, Stefan and Liertzer, Matthias and Persson, Emil and Burgd{\"o}rfer, Joachim and Lindroth, Eva},
	journal={Phys. Rev. A},
	volume={87},
	number={5},
	pages={053405},
	year={2013},
	publisher={APS},
	doi={10.1103/physreva.87.053405},
	url={https://doi.org/10.1103/physreva.87.053405}
}

@article{argenti2010,
	title={Ionization branching ratio control with a resonance attosecond clock},
	author={Argenti, Luca and Lindroth, Eva},
	journal={Phys. Rev. Lett.},
	volume={105},
	number={5},
	pages={053002},
	year={2010},
	publisher={APS},
	doi={10.1103/physrevlett.105.053002},
	url={https://doi.org/10.1103/physrevlett.105.053002}
}

@article{chu2010,
	title={Theory of ultrafast autoionization dynamics of {Fano} resonances},
	author={Chu, W-C and Lin, CD},
	journal={Phys. Rev. A},
	volume={82},
	number={5},
	pages={053415},
	year={2010},
	publisher={APS},
	doi={10.1103/physreva.82.053415},
	url={https://doi.org/10.1103/physreva.82.053415}
}

@article{mercouris2007,
	title={Time-dependent formation of the profile of the {He} {$2s2p\,^1P^o$} state excited by a short laser pulse},
	author={Mercouris, Th and Komninos, Yannis and Nicolaides, Cleanthes A},
	journal={Phys. Rev. A},
	volume={75},
	number={1},
	pages={013407},
	year={2007},
	publisher={APS},
	doi={10.1103/physreva.75.013407},
	url={https://doi.org/10.1103/physreva.75.013407}
}

@article{wickenhauser2005,
	title={Time resolved {Fano} resonances},
	author={Wickenhauser, Marlene and Burgd{\"o}rfer, Joachim and Krausz, Ferenc and Drescher, Markus},
	journal={Phys. Rev. Lett.},
	volume={94},
	number={2},
	pages={023002},
	year={2005},
	publisher={APS},
	doi={10.1103/physrevlett.94.023002},
	url={https://doi.org/10.1103/physrevlett.94.023002}
}

@article{Kaldun2016,
	author = {Kaldun, Andreas and Bl{\"a}ttermann, Alexander and Stoo{\ss}, Veit and Donsa, Stefan and Wei, Hui and Pazourek, Renate and Nagele, Stefan and Ott, Christian and Lin, Chii-Dong and Burgd{\"o}rfer, Joachim and T. Pfeifer },
	title = {Observing the ultrafast buildup of a {Fano} resonance in the time domain},
	journal = {Science},
	volume = {354},
	number = {6313},
	pages = {738-741},
	year = {2016},
	doi={10.1126/science.aah6972},
	url={https://doi.org/10.1126/science.aah6972}
}

@article{ott2013lorentz,
	title={{Lorentz} meets {Fano} in spectral line shapes: a universal phase and its laser control},
	author={Ott, Christian and Kaldun, Andreas and Raith, Philipp and Meyer, Kristina and Laux, Martin and Evers, J{\"o}rg and Keitel, Christoph H and Greene, Chris H and Pfeifer, Thomas},
	journal={Science},
	volume={340},
	number={6133},
	pages={716--720},
	year={2013},
	publisher={American Association for the Advancement of Science},
	doi={10.1126/science.1234407},
	url={https://doi.org/10.1126/science.1234407}
}

@article{xu2000scattering,
	title={Scattering-theory analysis of waveguide-resonator coupling},
	author={Xu, Yong and Li, Yi and Lee, Reginald K and Yariv, Amnon},
	journal={Phys. Rev. E},
	volume={62},
	number={5},
	pages={7389},
	year={2000},
	publisher={APS},
	doi={10.1103/physreve.62.7389},
	url={https://doi.org/10.1103/physreve.62.7389}
}

@article{luk2010,
	title={The {Fano} resonance in plasmonic nanostructures and metamaterials},
	author={Luk'Yanchuk, Boris and Zheludev, Nikolay I and Maier, Stefan A and Halas, Naomi J and Nordlander, Peter and Giessen, Harald and Chong, Chong Tow},
	journal={Nat. Mater.},
	volume={9},
	number={9},
	pages={707--715},
	year={2010},
	publisher={Nature Publishing Group UK London},
	doi={10.1038/nmat2810},
	url={https://doi.org/10.1038/nmat2810}
}

@article{miroshnichenko2010,
	title={{Fano} resonances in nanoscale structures},
	author={Miroshnichenko, Andrey E and Flach, Sergej and Kivshar, Yuri S},
	journal={Rev. Mod. Phys.},
	volume={82},
	number={3},
	pages={2257--2298},
	year={2010},
	publisher={APS},
	doi={10.1103/revmodphys.82.2257},
	url={https://doi.org/10.1103/revmodphys.82.2257}
}

@article{fano1961,
	title={Effects of configuration interaction on intensities and phase shifts},
	author={Fano, Ugo},
	journal={Phys. Rev.},
	volume={124},
	number={6},
	pages={1866},
	year={1961},
	publisher={APS},
	doi={10.1103/physrev.124.1866},
	url={https://doi.org/10.1103/physrev.124.1866}
}

@book{BetheSalpeter,
	author={Bethe, Hans A. and Salpeter, Edwin E.},
	title={Quantum Mechanics of One- and Two-Electron Atoms},
	publisher={Springer},
	address={Berlin},
	year={1957},
	doi={10.1007/978-3-662-12869-5},
	url={https://doi.org/10.1007/978-3-662-12869-5}
}

@article{Hoogenraad1998,
	author={Hoogenraad, J. H. and Noordam, L. D.},
	title={{Rydberg} atoms in far-infrared radiation fields: {I}. {Dipole} matrix elements of {H}, {Li}, and {Rb}},
	journal={Phys. Rev. A},
	volume={57},
	number={6},
	pages={4533--4541},
	year={1998},
	publisher={APS},
	doi={10.1103/physreva.57.4533},
	url={https://doi.org/10.1103/physreva.57.4533}
}

@article{LaneThomas1958,
	author={Lane, A. M. and Thomas, R. G.},
	title={R-matrix theory of nuclear reactions},
	journal={Rev. Mod. Phys.},
	volume={30},
	number={2},
	pages={257},
	year={1958},
	publisher={APS},
	doi={10.1103/revmodphys.30.257},
	url={https://doi.org/10.1103/revmodphys.30.257}
}

@article{FanoCooper1968,
	author={Fano, U. and Cooper, J. W.},
	title={Spectral distribution of atomic oscillator strengths},
	journal={Rev. Mod. Phys.},
	volume={40},
	number={3},
	pages={441},
	year={1968},
	publisher={APS},
	doi={10.1103/revmodphys.40.441},
	url={https://doi.org/10.1103/revmodphys.40.441}
}

@incollection{Starace1982,
	author={Starace, A. F.},
	title={Theory of atomic photoionization},
	booktitle={Handbuch der Physik},
	editor={Mehlhorn, W.},
	volume={31},
	pages={1--121},
	publisher={Springer},
	address={Berlin},
	year={1982},
	doi={10.1007/978-3-642-46453-9_1},
	url={https://doi.org/10.1007/978-3-642-46453-9_1}
}

@incollection{HuestisBerkowitz2011,
	author={Huestis, D. L. and Berkowitz, J.},
	title={Critical evaluation of the photoabsorption cross section of {CO$_2$} from 0.125 to 201.6 nm at room temperature},
	booktitle={Advances in Geosciences, Volume 25: Planetary Science},
	publisher={World Scientific},
	address={Singapore},
	pages={229--242},
	year={2011},
	doi={10.1142/9789814355377_0018},
	url={https://doi.org/10.1142/9789814355377_0018}
}

@article{Henning1932,
	author={Henning, H. J.},
	title={{Die Absorptionsspektren von Kohlendioxyd, Kohlenmonoxyd und Wasserdampf im Gebiet von 600--900 {\AA}E}},
	journal={Ann. Phys. (Leipzig)},
	volume={405},
	number={5},
	pages={599--620},
	year={1932},
	doi={10.1002/andp.19324050505},
	url={https://doi.org/10.1002/andp.19324050505}
}

@article{TanakaJursaLeBlanc1960,
	author={Tanaka, Y. and Jursa, A. S. and LeBlanc, F. J.},
	title={Higher ionization potentials of linear triatomic molecules. {I}. {CO$_2$}},
	journal={J. Chem. Phys.},
	volume={32},
	number={4},
	pages={1199--1205},
	year={1960},
	doi={10.1063/1.1730874},
	url={https://doi.org/10.1063/1.1730874}
}

@article{Plemelj1908,
	author={Plemelj, J.},
	title={{Ein Erg{\"a}nzungssatz zur Cauchyschen Integraldarstellung analytischer Funktionen, Randwerte betreffend}},
	journal={Monatsh. Math. Phys.},
	volume={19},
	pages={205--210},
	year={1908},
	doi={10.1007/BF01736696},
	url={https://doi.org/10.1007/BF01736696}
}

@book{Sakurai2020,
	author={Sakurai, J. J. and Napolitano, J.},
	title={Modern Quantum Mechanics},
	edition={3},
	publisher={Cambridge University Press},
	address={Cambridge},
	year={2020},
	doi={10.1017/9781108587280},
	url={https://doi.org/10.1017/9781108587280}
}

@article{SignorellMerkt1997,
	title={General symmetry selection rules for the photoionization of polyatomic molecules},
	author={Signorell, Ruth and Merkt, Fr{\'e}d{\'e}ric},
	journal={Mol. Phys.},
	volume={92},
	number={5},
	pages={793--804},
	year={1997},
	doi={10.1080/002689797169745},
	url={https://doi.org/10.1080/002689797169745}
}

@article{Heays2017,
	title={Photodissociation and photoionisation of atoms and molecules of astrophysical interest},
	author={Heays, A. N. and Bosman, A. D. and van Dishoeck, E. F.},
	journal={Astron. Astrophys.},
	volume={602},
	pages={A105},
	year={2017},
	doi={10.1051/0004-6361/201628742},
	url={https://doi.org/10.1051/0004-6361/201628742}
}

@article{Harris1978,
	title={On the use of windows for harmonic analysis with the discrete {Fourier} transform},
	author={Harris, Fredric J.},
	journal={Proc. IEEE},
	volume={66},
	number={1},
	pages={51--83},
	year={1978},
	doi={10.1109/PROC.1978.10837}
}

@article{Lam2022,
	title = {Alignment dependence of photoelectron angular distributions in the few-photon ionization of molecules by ultraviolet pulses},
	author = {Lam, Huynh Van Sa and Wangjam, Tomthin Nganba and Kumarappan, Vinod},
	journal = {Phys. Rev. A},
	volume = {105},
	pages = {053109},
	year = {2022},
	doi = {10.1103/PhysRevA.105.053109}
}

@article{Wangjam2021,
	title = {Strong-field ionization of the triplet ground state of $\text{O}_2$},
	author = {Wangjam, Tomthin Nganba and Lam, Huynh Van Sa and Kumarappan, Vinod},
	journal = {Phys. Rev. A},
	volume = {104},
	pages = {043112},
	year = {2021},
	doi = {10.1103/PhysRevA.104.043112}
}

@article{Li2016,
	title = {Retrieving plasmonic near-field information: A quantum-mechanical model for streaking photoelectron spectroscopy of gold nanospheres},
	author = {Li, Jianxiong and Saydanzad, Erfan and Thumm, Uwe},
	journal = {Phys. Rev. A},
	volume = {94},
	pages = {051401(R)},
	year = {2016},
	doi = {10.1103/PhysRevA.94.051401}
}

@article{Saydanzad2017,
	title = {Characterization of induced nanoplasmonic fields in time-resolved photoemission: A classical trajectory approach applied to gold nanospheres},
	author = {Saydanzad, Erfan and Li, Jianxiong and Thumm, Uwe},
	journal = {Phys. Rev. A},
	volume = {95},
	pages = {053406},
	year = {2017},
	doi = {10.1103/PhysRevA.95.053406}
}

@article{Li2018PRL,
	title = {Imaging Plasmonic Fields with Atomic Spatiotemporal Resolution},
	author = {Li, Jianxiong and Saydanzad, Erfan and Thumm, Uwe},
	journal = {Phys. Rev. Lett.},
	volume = {120},
	pages = {223903},
	year = {2018},
	doi = {10.1103/PhysRevLett.120.223903}
}

@article{Saydanzad2018,
	title = {Spatiotemporal imaging of plasmonic fields near nanoparticles below the diffraction limit},
	author = {Saydanzad, Erfan and Li, Jianxiong and Thumm, Uwe},
	journal = {Phys. Rev. A},
	volume = {98},
	pages = {063422},
	year = {2018},
	doi = {10.1103/PhysRevA.98.063422}
}

@article{Powell2022,
	title = {Strong-Field Control of Plasmonic Properties in Core--Shell Nanoparticles},
	author = {Powell, Jeffrey A. and Li, Jianxiong and Summers, Adam and Robatjazi, Seyyed Javad and Davino, Michael and Rupp, Philipp and Saydanzad, Erfan and Sorensen, Christopher M. and Rolles, Daniel and Kling, Matthias F. and Trallero, Carlos and Thumm, Uwe and Rudenko, Artem},
	journal = {ACS Photonics},
	volume = {9},
	number = {11},
	pages = {3515--3521},
	year = {2022},
	doi = {10.1021/acsphotonics.2c00663}
}

@article{Saydanzad2022,
	title = {Strong-field ionization of plasmonic nanoparticles},
	author = {Saydanzad, Erfan and Li, Jianxiong and Thumm, Uwe},
	journal = {Phys. Rev. A},
	volume = {106},
	pages = {033103},
	year = {2022},
	doi = {10.1103/PhysRevA.106.033103}
}

@article{Saydanzad2023,
	title = {Enhanced cutoff energies for direct and rescattered strong-field photoelectron emission of plasmonic nanoparticles},
	author = {Saydanzad, Erfan and Powell, Jeffrey and Summers, Adam and Robatjazi, Seyyed Javad and Trallero-Herrero, Carlos and Kling, Matthias F. and Rudenko, Artem and Thumm, Uwe},
	journal = {Nanophotonics},
	volume = {12},
	number = {10},
	pages = {1931--1942},
	year = {2023},
	doi = {10.1515/nanoph-2023-0120}
}

@article{Saydanzad2025,
	title = {Generation of fast photoelectrons in strong-field emission from metal nanoparticles},
	author = {Saydanzad, Erfan and Powell, Jeffrey and Renner, Tim and Summers, Adam and Rolles, Daniel and Trallero-Herrero, Carlos and Kling, Matthias F. and Rudenko, Artem and Thumm, Uwe},
	journal = {Nanophotonics},
	volume = {14},
	number = {9},
	pages = {1355--1364},
	year = {2025},
	doi = {10.1515/nanoph-2024-0719}
}
